\documentclass[12pt]{article}
\usepackage[margin=1.0in]{geometry}
\usepackage[utf8]{inputenc}
\title{Balanced Stochastic Block Model for Community Detection in Signed Networks}
\date{}

\usepackage[style=authoryear,maxcitenames=1,maxbibnames=99,backend=biber,alldates=comp, giveninits=true, uniquename=false, uniquelist=false,dashed=false]{biblatex}
\DeclareNameAlias{sortname}{family-given}
\AtEveryBibitem{%
  \clearfield{month}%
}

\usepackage{amsfonts,amsmath,amsthm,amssymb,xcolor,graphicx,bm,bbm}
\usepackage{enumerate}
\usepackage{physics}
\usepackage{enumitem}
\usepackage{caption}
\usepackage{multirow}
\usepackage{enumitem}
\usepackage{booktabs}
\usepackage{mathtools}
\usepackage{setspace}
\usepackage{graphicx}
\usepackage{subcaption}
\usepackage{booktabs}
\usepackage{array} 
\usepackage{ragged2e} 
\usepackage{float} 
\usepackage{setspace}
\usepackage[flushleft]{threeparttable} 
\usepackage{booktabs,caption}

\usepackage{soul}
\usepackage[ruled,linesnumbered]{algorithm2e}
\usepackage{algpseudocode} 
\definecolor{darkpowderblue}{rgb}{0.0, 0.2, 0.6}
\definecolor{goldenpoppy}{rgb}{0.99, 0.76, 0.0}
\definecolor{cardinal}{rgb}{0.77, 0.12, 0.23}
\usepackage{hyperref}
\hypersetup{
    colorlinks=true,
    linkcolor=cardinal,
    citecolor=darkpowderblue,      
    urlcolor=cyan,
}

\newcommand{\mc}[1]{\mathcal{#1}}
\newcommand{\mb}[1]{\mbox{\textbf{#1}}}

\DeclareMathOperator*{\argmax}{arg\,max}

\newtheorem{theorem}{Theorem}
\newtheorem*{theorem*}{Theorem}

\newtheorem{remark}{Remark}
\newtheorem{proposition}{Proposition}

\newtheorem{corollary}{Corollary}[theorem]

\allowdisplaybreaks

\author{
  Yichao Chen\footnote{These authors contributed equally to this work.} \\
  Department of Statistics, University of Michigan \\[1ex]  
  Weijing Tang\footnotemark[1] \\
   Department of Statistics \& Data Science, Carnegie Mellon University\\[1ex]
  Ji Zhu \\
  Department of Statistics, University of Michigan
}

\begin{document}
\maketitle

\begin{abstract}
     Community detection, discovering the underlying communities within a network from observed connections, is a fundamental problem in network analysis, yet it remains underexplored for \textit{signed networks}. In signed networks, both edge connection patterns and edge signs are informative, and structural balance theory (e.g., triangles aligned with ``the enemy of my enemy is my friend'' and ``the friend of my friend is my friend'' are more prevalent) provides a global higher-order principle that guides community formation.  We propose a Balanced Stochastic Block Model (BSBM), which incorporates balance theory into the network generating process such that balanced triangles are more likely to occur. We develop a fast profile pseudo-likelihood estimation algorithm with provable convergence and establish that our estimator achieves strong consistency under weaker signal conditions than methods for the binary SBM that rely solely on edge connectivity. Extensive simulation studies and two real-world signed networks demonstrate strong empirical performance.
\end{abstract}

\section{Introduction}
In network analysis, \textit{communities} are defined as clusters of nodes whose members share similar connection patterns with others. Community detection, discovering such latent clusters from an observed network, is a fundamental problem that has received extensive attention.  
Many methods for community detection are based on probabilistic network models, including the stochastic block model (SBM)~\parencite{holland1983stochastic, nowicki2001estimation}, degree-corrected SBM \parencite{karrer2011stochastic}, latent factor model \parencite{handcock2007model,hoff2007modeling}, and mixed-membership SBM for overlapping community detection \parencite{airoldi2008mixed}.  Other methods formulate community detection as an optimization problem, which maximizes criteria that quantify the strength of community structure or their spectral approximations, including normalized cuts \parencite{shi2000normalized}, modularity \parencite{newman2004finding,newman2006modularity}, and variants of spectral clustering \parencite{ng2001spectral}. 
These methods rely solely on the edge-connectivity information in binary networks for community detection.

In many applications, however, networks contain not only information about whether a connection exists but also the type of the connection. In \textit{signed networks}, each edge takes either a positive (e.g., friendship, trust, agreement, positive correlation) or negative (e.g., hostility, distrust, disagreement, negative correlation) sign. Such signed networks are common in diverse fields, examples include social network~\parencite{heider1946attitudes,leskovec2010signed}, international relations~\parencite{doreian1996partitioning,doreian2015structural,tang2025population}, and biological network~\parencite{vinayagam2014integrating, morabito2023hdwgcna}. 
Incorporating edge-sign information in signed networks allows for the identification of community structures that are not captured by edge-connectivity patterns alone. 

To this end, numerous algorithms have been proposed for community detection in signed networks~\parencite{doreian1996partitioning,bansal2004correlation,yang2007community,chiang2012scalable,li2014comparative, kunegis2010spectral}, among which many extend classical criteria such as normalized cuts and modularity that were originally designed for binary (unsigned) networks to incorporate edge signs. 
These extensions aggregate local pairwise sign signals into a partition objective, where positive edges encourage placing the connected nodes in the same community while negative edges encourage assigning them to different communities. 
However, methods based on local pairwise information alone overlooks a unique feature of signed networks: positive and negative edges interact through higher-order patterns. An important theory in social psychology for understanding such interactions is
\textit{structural balance theory} \parencite{harary1953notion}. The theory characterizes signed triangles (e.g., three nodes connected to each other) as either \textit{balanced} if the product of their three edge
signs is positive, or \textit{unbalanced} otherwise. 
In particular, balanced triangles are consistent with the proverbs ``the enemy of my enemy is my friend" and ``the friend of my friend is my friend".
The balance theory suggests that balanced triangles are more prevalent than unbalanced ones in signed networks. This pattern has been empirically observed in numerous real-world signed networks, including social and biological networks \parencite{facchetti2011computing, allahyari2022structure,aref2018measuring}. 
Balance theory provides a global higher-order principle for community detection. Beyond local pairwise information, it suggests that communities should be formed to minimize the occurrence of unbalanced triangles across the network.  To incorporate the structural balance, one line of work uses low-rank matrix completion algorithms for community detection and sign prediction~\parencite{hsieh2012low,chiang2014prediction}. However, they treat non-edges as missing entries and thereby rely solely on edge-sign information. 

Probabilistic model-based approaches for signed networks, in contrast, have been  less explored~\parencite{Vu2013-up,chen2014overlapping,jiang2015stochastic,zhang2022signed,li2023ssbm,tang2025population, pensky2025signed}. Among them, \textcite{jiang2015stochastic} transformed a signed network into a two-layer network, where one layer represents the presence of positive edges and the other represents the presence of negative edges. This approach, however, ignores the mutual exclusivity between positive and negative edges on the same node pair. \textcite{Vu2013-up} developed an exponential random network model for discrete-valued networks, which can be applied to signed networks, and \textcite{li2023ssbm} proposed a signed SBM, where each edge follows a multinomial distribution. More recently, \textcite{pensky2025signed} proposed a variant of the generalized random dot product graph model for signed networks. 
However, all of them do not incorporate the balance theory for community detection that is the focus of our work.  \textcite{zhang2022signed} introduced a latent space model for joint community and anomaly detection. Their method models signed edges as ordinal variables in $\{-1, 0, 1\}$, interpreting a non-edge as a neutral status between positive and negative. However, this assumption may not hold in practice, especially for large sparse networks. For example, two individuals with opposing political views may have no observed connection not because they hold neutral attitudes, but simply because their social circles do not intersect. 
In comparison, \textcite{tang2025population} treated signed edges as categorical variables with three levels and introduced a stochastic notion of \textit{population-level balance}. They established sufficient conditions under which signed network generated from a broad class of latent variable models are inherently population-level balanced. While their sufficient condition offers insights that could guide community detection, it remains unclear how population-level balance can be effectively leveraged to perform community detection. 

To bridge this gap, we develop a novel probabilistic model-based method guided by balance theory for community detection in signed networks, which integrates both the edge-connectivity and edge-sign information. Our contributions consist of three parts: 
\begin{enumerate}
\item We propose a balanced stochastic block model (BSBM) for signed networks, which treats signed edges as categorical variables. Unlike existing SBMs for signed networks, our model introduces a hierarchical \textit{meta-group} structure in the sign-generation process to incorporate population-level balance. Specifically, communities are mapped to one of two meta-groups, where edges are more likely to be negative between meta-groups and positive within meta-groups. Therefore, negative edges tend to occur between communities rather than within them. This hierarchical design naturally incorporates structural balance theory to guide community detection.
\item  While the hierarchical design effectively incorporates balance theory, it also makes the model estimation, already a well-known challenge in fitting block models, even more complicated. To address this, we develop a fast profile-pseudo likelihood estimation method for the BSBM with a convergence guarantee. Our method builds on the idea of decoupling node memberships for rows and columns~\parencite{Wang2021-pp}, but differs in several aspects. In contrast to their binary SBM, the hierarchical structure in the BSBM couples each community with a higher-level meta-group, leading to a binary quadratic optimization subproblem without closed-form updates. We solve this subproblem exactly when the number of communities is small and approximately via semi-definite programming when the number is large.
\item  We also establish the strong consistency of community membership estimation in scenarios where strong consistency is not achievable using edge-connection information alone. Note that in binary SBM, strong consistency requires a sufficiently large gap between the within- and between-community connection probabilities. Our method can achieve strong consistency under weaker conditions on this gap by leveraging both edge-connectivity and edge-sign information, provided that the latter offers sufficient separation between communities.
\end{enumerate}

The rest of this paper is organized as follows. Section \ref{sec:model} introduces our proposed model and presents an illustrative example demonstrating the advantage of incorporating edge-sign information.   Section \ref{sec:estimation} provides the details of our estimation method and algorithm with a convergence guarantee. Section \ref{sec:theory} establishes the strong consistency. Sections \ref{sec:simulation} and \ref{sec:realdata} demonstrate the performance of our method through extensive simulation studies and two real-world data applications. Section \ref{sec:conclusion} concludes our work.  

\section{Balanced Stochastic Block Model} 
\label{sec:model}
The stochastic block model (SBM) is a canonical probabilistic framework for studying community detection. In this section, we introduce an SBM for signed networks and impose a structure to incorporate the balance theory. 

Denote an undirected signed network by a symmetric adjacency matrix $\mb{A}=[A_{ij}]_{1\le i, j \le n} \in \{-1,0,1\}^{n\times n}$, where $A_{ij}=1$ if there is a positive edge between node $i$ and node $j$, $A_{ij}=-1$ if there is a negative edge, and $A_{ij}=0$ if there is no edge. Assume there are $K$ communities and let $\mb{z}=(z_1, \cdots, z_n)\in [K]^n$ denote the community memberships of $n$ nodes, with $z_i$ i.i.d.\ drawn from a categorical distribution ${\rm Cat}(\boldsymbol{\pi})$ with probabilities $\boldsymbol{\pi}=(\pi_1, \cdots, \pi_K)$ satisfying $\sum_{\ell=1}^K\pi_\ell=1$. We consider an SBM for signed networks, where edges are generated independently given the node community memberships. Specifically, an edge between nodes $i$ and $j$ forms with probability
\begin{align}
	Pr(|A_{ij}|=1 \mid \mb{z})&=P_{z_i, z_j}, \label{eq: edge sbm}
\end{align}
and, given that the edge exists, it is positive with probability 
\begin{align}
    Pr(A_{ij}=1 \mid |A_{ij}|=1, \mb{z})=Q_{z_i, z_j}. \label{sign sbm}
\end{align}
Here, $\mb{P}=[P_{\ell \ell'}]_{1\le \ell, \ell' \le K}\in [0,1]^{K\times K}$ and $\mb{Q}=[Q_{\ell \ell'}]_{1\le \ell, \ell' \le K}\in [0,1]^{K\times K}$ are  symmetric matrices. The entry $P_{\ell\ell'}$ specifies the probability of formation of an edge between a node in the community $\ell$ and a node in the community $\ell'$, while $Q_{\ell\ell'}$ specifies the probability that such an edge is positive. 
This model formulation is equivalent to modeling each edge as a three-category multinomial variable as in \textcite{li2023ssbm}, since a multinomial with three categories can be parameterized by two probabilities. Our formulation separates edge formation and sign assignment, which makes it easier to interpret and incorporate structural balance theory.

\begin{figure}[!h]
    \centering
    \includegraphics[width=0.5\linewidth]{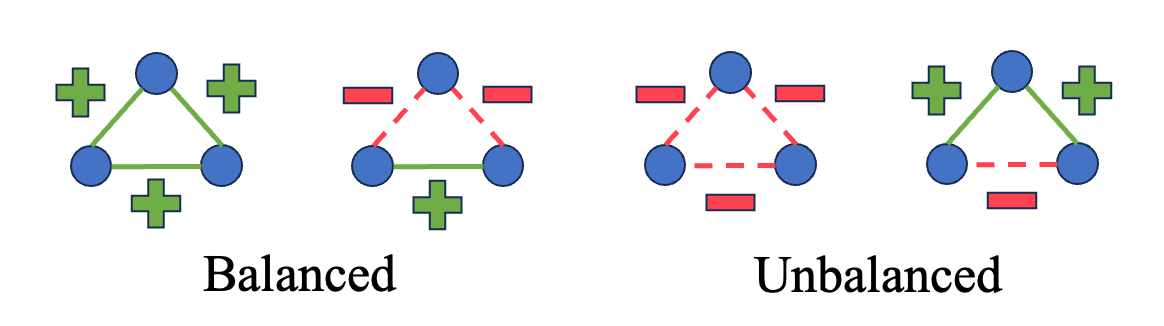}
    \caption{The left two are balanced triangles and the right two are unbalanced triangles}
    \label{fig:demostration of population-balance}
\end{figure}

\begin{figure}[!h]
 \begin{center}
 \includegraphics[width=0.5\linewidth]{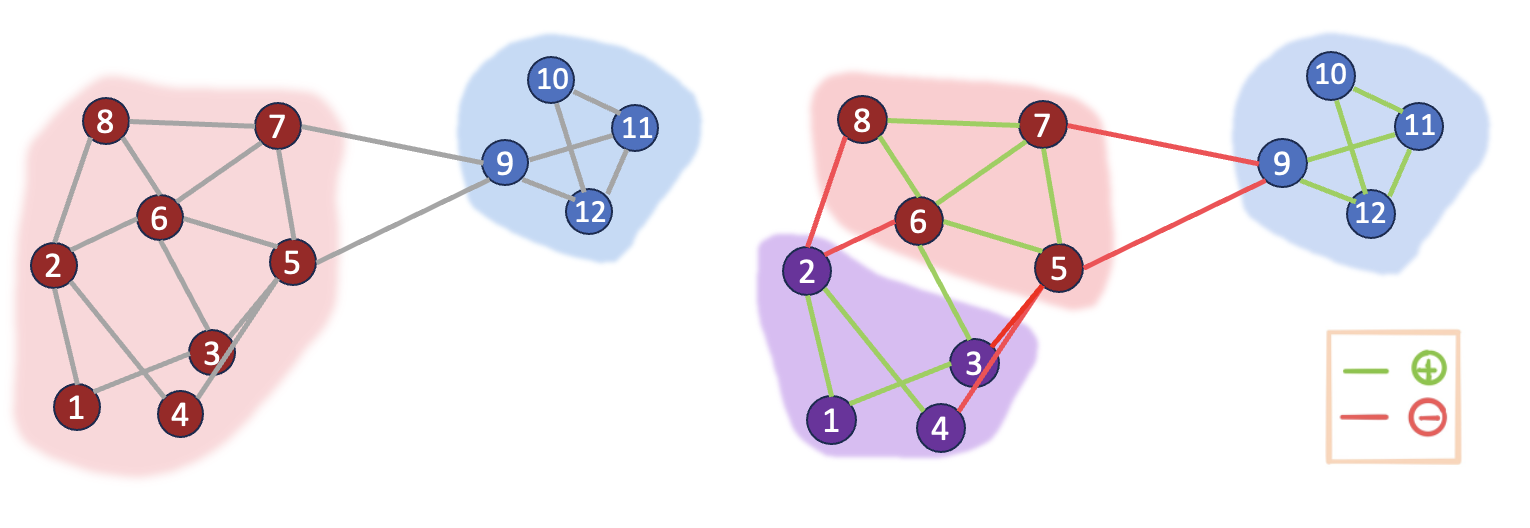}
 \end{center}
 \caption{The network contains three communities: nodes labeled 1 to 4, nodes labeled 5 to 8, and nodes labeled 9 to 12. Nodes 1–4 and 9–12 belong to meta-group $G_1$, and nodes 5–8 belong to meta-group $G_2$. \emph{Left:} Community detection under a vanilla SBM using unsigned connectivity (grey edges).  \emph{Right:} Community detection under the BSBM using signed edges (positive in green and negative in red). }
 \label{fig: bsbm}
\end{figure}

We impose additional structure on $\mb{Q}$ to incorporate \textit{population-level balance}, a stochastic notion of balance theory that accounts for the noise in the observed data~\parencite{tang2025population}. 
A signed network is said to be population-level balanced if 
 $\mathbb{E}(A_{ij}A_{jk}A_{ki}\mid |A_{ij}A_{jk}A_{ki}|=1)>0$
 for any three different nodes $1\le i,j,k\le n$. 
Intuitively, this definition requires that, in expectation, triangles are more likely to have a positive product of edge signs. Recall that balanced triangles correspond to triangles with a positive product, while unbalanced triangles have a negative product (see Figure~\ref{fig:demostration of population-balance} for illustration). Thus, population-level balance suggests that balanced triangles are more prevalent than unbalanced ones. 
Following \textcite{tang2025population}, one sufficient condition for achieving population-level balance is \textit{latent-space separability}, where the latent space can be partitioned into two regions such that edges connecting nodes within the same region are more likely to have positive signs, while edges across regions are more likely to have negative signs. Motivated by this insight, we parameterize the sign probability matrix $\mathbf{Q}$ as 
\begin{equation}
Pr(A_{ij}=1 \mid |A_{ij}|=1, \mb{z})= Q_{z_i z_j} =  \frac{1 +\eta_{z_i z_j} \nu\left(z_i\right) \nu\left(z_j\right)}{2},
\label{Q structure}
\end{equation}
where $\eta_{\ell \ell^{\prime}}=\eta_{\ell^{\prime} \ell} \in[0,1]$ for $1 \leq \ell, \ell^{\prime} \leq K$ and $\nu:[K] \rightarrow\{-1,1\}$ maps each community to one of two groups. 
The mapping $\nu(\cdot)$ thus induces a hierarchical partition of $K$ communities into two \textit{meta-groups}: $G_1=\{\ell \in [K]: \nu(\ell)=1\}$ and $G_2=\{\ell \in [K]: \nu(\ell)=-1\}$. Under this structure, pairs of nodes whose communities share the same meta-group label, i.e., $\nu(\ell)\nu(\ell') = 1$, are more likely to have positive signs, whereas pairs across different meta-groups are more likely to have negative signs.
We refer to this structured SBM as the \textit{Balanced Stochastic Block Model} (BSBM). The following proposition formalizes that this hierarchical structure induced by $\nu(\cdot)$ guarantees population-level balance.
\begin{proposition} For $\mb{A} \sim BSBM$,  $\mathbb{E}(A_{ij}A_{jk}A_{ki}\mid |A_{ij}A_{jk}A_{ki}|=1)>0$ for any $1\le i<j<k\le n$. 
\label{Proposition 1}
\end{proposition} 
The proof of Proposition \ref{Proposition 1} is in Appendix. Our proposed BSBM integrates balance theory into the generation process of signed networks, which leads to globally coherent sign patterns. As a result, BSBM can reveal community structure when edge connectivity alone provides weak signals. 
In particular, under BSBM, a negative edge between two nodes suggests that they likely belong to different meta-groups, and hence to different communities. 
Figure~\ref{fig: bsbm} illustrates this effect: in the left panel, communities 1 and 2 are difficult to distinguish using only unsigned edge connectivity information, whereas the right panel that incorporates edge-sign information enables a clear separation between them. 

\section{Maximum Profile-Pseudo Likelihood Estimation}
\label{sec:estimation}
Fitting the likelihood function of an SBM has been known as a non-trivial task, as maximizing the likelihood for the observed data requires summing over all possible community label assignments, which is an NP-hard problem in general. 
Specifically, the likelihood of complete data for $(\mb{A},\mb{z})$ under BSBM is given by
\begin{align*}
\mathcal{L}(\boldsymbol{\pi}, \mb{P}, \mb{Q}; \mb{A},\mb{z}) = 
\prod_{i \in [n]}  \pi_{z_i} 
\prod_{1 \leq i < j \leq n} \left( Q_{z_i z_j}^{B_{ij}} \left(1 - Q_{z_i z_j}\right)^{1- B_{ij}} P_{z_i z_j}\right)^{|A_{ij}|} \left( 1 - P_{z_i z_j}\right)^{1 - |A_{ij}|},
\end{align*}
where $B_{ij} = (1 + A_{ij})/2$, and $Q_{\ell \ell'} =  \frac{1 +\eta_{\ell \ell'} \nu\left(\ell\right) \nu\left(\ell'\right)}{2}.$ Here, $B_{ij} = 1$ corresponds to a  positive edge and $B_{ij} = 0$ corresponds to a negative edge. 
Optimizing the likelihood of observed data $\mb{A}$ requires marginalizing the complete-data likelihood over all possible $\mb{z}$, which is computationally intractable.
Moreover, the constraint imposed on the sign probability matrix $\mb{Q}$ introduces additional dependencies between community labels and parameters, which makes optimization more challenging. 

To address these challenges, we develop a maximum profile-pseudo likelihood estimation method, which decouples the community labels associated with the rows and columns in the likelihood function. This approach is inspired by the recent advance in fast likelihood optimization for a binary (unsigned) SBM proposed in \textcite{Wang2021-pp}. We extend this decoupling strategy to the signed networks and adapt it to accommodate the additional hierarchical constraint on $\mb{Q}$ required for population-level balance under BSBM.

Let $\boldsymbol{\Omega}=(\boldsymbol{\pi}, \mb{P}, \mb{Q})$ denote the model parameters. We define the parameter space as 
\begin{align*}
    \boldsymbol \Theta = \Bigg\{ (\boldsymbol{\pi}, \mb{P}, \mb{Q}): & \boldsymbol{\pi}=(\pi_1, \cdots, \pi_K)\in [0,1]^K \text{ with }  \sum_{\ell=1}^K\pi_\ell=1, \  \mb{P} \in [0,1]^{K \times K}, \\ 
    & \mb{Q} = (1+  \boldsymbol{\eta} \circ \boldsymbol{\nu} \boldsymbol{\nu}^\top )/2 \ \text{ with }\ \boldsymbol{\eta}\in [0,1]^{K \times K}\  \text{ and }\ \boldsymbol{\nu} \in \{\pm1\}^K   \Bigg\}.
\end{align*}
Let $\boldsymbol{e}=\left(e_1, e_2, \ldots, e_n\right) \in [K]^n$ denote the column labeling vector. We treat the row labels $\mb{z}$ as random latent variables and the column labels~$\boldsymbol{e}$ as fixed but unknown model parameters. This decoupling allows us to obtain a tractable expression when optimizing with respect to the column labels $\boldsymbol{e}$ and other nuisance model parameters $\boldsymbol{\Omega}$. 
When treating the column label $e_j$ as a fixed parameter, each row $A_{i\star}=(A_{i1}, \cdots, A_{in})$ i.i.d.\ follows a mixture of $K$ categorical distributions, where $z_i \sim {\rm Cat}(\boldsymbol{\pi})$ and conditional on $z_i = \ell$, for $1\le j \le n$,
\[
A_{ij}  =
\begin{cases}
1, & \text{with probability } P_{\ell e_j} Q_{\ell e_j},\\[3pt]
-1, & \text{with probability } P_{\ell e_j} (1 - Q_{\ell e_j}),\\[3pt]
0, & \text{with probability } 1 - P_{\ell e_j}.
\end{cases}
\]
We define the corresponding log-likelihood of i.i.d.\ observations $\{A_{i\star}\}_{i =1}^n$ under this mixture model as the \textit{log-pseudo-likelihood function}:
\begin{align}
\mc{L}_{\mathrm{PL}}\left(\boldsymbol{\Omega}, \boldsymbol{e} ; \mb{A}\right)=\sum_{i=1}^n \log\left\{\sum_{\ell=1}^K \pi_\ell \prod_{j=1}^n \left(Q_{\ell{e_j}}^{B_{i j}}\left(1-Q_{\ell{e_j}}\right)^{1-B_{i j}}P_{\ell{e_j}}\right)^{|A_{ij}|} \left(1-P_{\ell{e_j}}\right)^{1-|A_{ij}|}\right\},
\label{eq:log-pseudo-likelihood}
\end{align}
which serves as an approximation to the original likelihood under BSBM. 

To optimize this pseudo-likelihood, we consider an iterative optimization algorithm that alternates between updating $\boldsymbol{e}$ and $\boldsymbol{\Omega}$. 
In each iteration, given the current estimate $\hat{\boldsymbol{e}}$, we first profile out the nuisance model parameters $\boldsymbol{\Omega}$ by solving $\widehat{\boldsymbol{\Omega}}=\argmax _{\boldsymbol{\Omega} \in  \boldsymbol \Theta }\mc{L}_{\mathrm{PL}}\left(\boldsymbol{\Omega}, \hat{\boldsymbol{e}}; \mb{A}\right)$. We then maximize the profile-pseudo likelihood with respect to $\boldsymbol{e}$, i.e., $\max_{\boldsymbol{e}}\mc{L}_{\mathrm{PL}}(\widehat{\boldsymbol{\Omega}}, \boldsymbol{e}; \mb{A})$. We provide an overview of our algorithm below, with detailed steps summarized in Algorithm~\ref{alg:profile_pseudo_likelihood}.
\begin{itemize}
	\item Given the current estimate $\hat{\boldsymbol{e}}$, we solve $\widehat{\boldsymbol{\Omega}}=\argmax _{\boldsymbol{\Omega} \in  \boldsymbol \Theta}\mc{L}_{\mathrm{PL}}\left(\boldsymbol{\Omega}, \hat{\boldsymbol{e}}; \mb{A}\right)$ via an expectation-maximization (EM) algorithm. In the E-step, the posterior probabilities of the latent row labels have closed-form expressions. In the M-step, the update of $(\boldsymbol{\pi},  \boldsymbol{P})$ also has closed forms, while the update of $\mb{Q}$ involves solving a binary quadratic optimization problem due to the constraint in the parameter space. 
	\item Given the current estimate $\widehat{\boldsymbol{\Omega}}$, we update the column labels $\hat{\boldsymbol{e}}$ by maximizing the expected log-pseudo-likelihood for complete data. 
    Instead of optimizing jointly over possible label assignments for all nodes, we adopt a fast local updating rule following \textcite{Wang2021-pp}, which guarantees a non-negative increment in the pseudo-likelihood after each iteration. 
\end{itemize}

\begin{algorithm}[H]
\caption{Profile-Pseudo Likelihood Maximization Algorithm}
\label{alg:profile_pseudo_likelihood}
\DontPrintSemicolon 

Initialize \( \boldsymbol e^{(0)} \) using spectral clustering with perturbations.\;
Calculate $\boldsymbol\Omega^{(0)}=(\boldsymbol{\pi}^{(0)},  \mathbf{P}^{(0)}, \mathbf{Q}^{(0)})$: for \( 1 \leq \ell \leq K \),\;
$\pi_\ell^{(0)} = \frac{1}{n} \sum_{i=1}^n I(e_i^{(0)} = \ell),$\;
$P_{\ell \ell'}^{(0)} = \frac{\sum_{i=1}^n \sum_{j=1}^n |A_{ij}|I(e_i^{(0)}=\ell)I(e_j^{(0)}=\ell')}{\sum_{i=1}^n \sum_{j=1}^n I(e_i^{(0)}=\ell)I(e_j^{(0)}=\ell')},$\;
$Q_{\ell\ell'}^{(0)} = \frac{\sum_{i=1}^n \sum_{j=1}^n I(A_{ij} = 1)I(e_i^{(0)}=\ell)I(e_j^{(0)}=\ell')}{\sum_{i=1}^n \sum_{j=1}^n |A_{ij}| I(e_i^{(0)}=\ell)I(e_j^{(0)}=\ell')}.$\;

Initialize $\boldsymbol \Omega^{(0,0)}=(\boldsymbol{\pi}^{(0,0)},  \mathbf{P}^{(0,0)}, \mathbf{Q}^{(0,0)}) = (\boldsymbol{\pi}^{(0)},  \mathbf{P}^{(0)}, \mathbf{Q}^{(0)}),$\;
\Repeat{the profile-pseudo likelihood converges}{
    \Repeat{the EM algorithm converges}{
        E-step: compute \( \tau_{i\ell}^{(s,t+1)} \) using \eqref{tau} for \( 1 \leq \ell \leq K \) and \( 1 \leq i \leq n \),\;
        M-step: compute \(\pi_k^{(s,t+1)} \), \( P_{\ell \ell'}^{(s,t+1)} \), and \( Q_{\ell \ell'}^{(s,t+1)}\)
        using \eqref{pi} - \eqref{Q} for \( 1 \leq \ell, \ell' \leq K \).\;
    }
    Set \( \boldsymbol\Omega^{(s+1)} \) and \( \{\tau_{i\ell}^{(s+1)}\} \) as the final EM update.\;
    Given \( \boldsymbol\Omega^{(s+1)} \) and \( \{\tau_{i\ell}^{(s+1)}\} \), update \( e_j^{(s+1)} \) using \eqref{e}. \;
}
\end{algorithm}

Our algorithm consists of two layers of iterations. The \textit{outer iterations} alternate between updating the column labels $\boldsymbol{e}$ and the model parameters $\boldsymbol{\Omega}$. Within each outer iteration, the model parameters $\boldsymbol{\Omega}$ are updated using an EM procedure, whose iterations are referred to as the \textit{inner iterations}.
The detailed updating rules for both layers of iterations are provided below.

\paragraph{Inner Iterations of EM Algorithm}
After the $s$-th outer iteration and the $t$-th inner iteration, we have the current estimates $\boldsymbol{e}^{(s)}$ and $\boldsymbol{\Omega}^{(s,t)}=(\boldsymbol{\pi}^{(s,t)}, \mb{P}^{(s,t)}, \mb{Q}^{(s,t)})$.  In the $(t+1)$-th inner iteration, the E-step and M-step proceed as follows. 
In the E-step, we compute the posterior probability that node $i$ belongs to community $\ell$ given the current estimate $\boldsymbol{\Omega}^{(s,t)}$ and the observed signed network, i.e.,  $\tau_{i \ell}^{(s, t+1)} = P(z_i = \ell \mid A_{i \star}, \boldsymbol{\Omega}^{(s,t)})$. This posterior probability has the following closed-form expression: for $1 \leq i \leq n$ and $1 \leq \ell \leq K$,
\begin{gather}
    \begin{aligned}
        \tau_{i \ell}^{(s, t+1)}=\frac{\pi_\ell^{(s, t)} \prod_{j=1}^n    \left(   \left\{Q_{\ell e_j^{(s)}}^{(s, t)}\right\}^{B_{i j}}\left\{1-Q_{\ell e_j^{(s)}}^{(s, t)}\right\}^{1-B_{i j}} P_{\ell e_j^{(s)}}^{(s, t)}\right)^{|A_{ij}|} \left(1 - P_{\ell e_j^{(s)}}^{(s, t)}\right)^{1 - |A_{ij}|}}{\sum_{\ell=1}^K \pi_\ell^{(s, t)} \prod_{j=1}^n \left(   \left\{Q_{\ell e_j^{(s)}}^{(s, t)}\right\}^{B_{i j}}\left\{1-Q_{\ell e_j^{(s)}}^{(s, t)}\right\}^{1-B_{i j}} P_{\ell e_j^{(s)}}^{(s, t)}\right)^{|A_{ij}|} \left(1 - P_{\ell e_j^{(s)}}^{(s, t)}\right)^{1 - |A_{ij}|}}.
    \end{aligned}
    \label{tau}
 \end{gather} 
In the M-step, we update $\boldsymbol{\Omega}^{(s, t+1)}$ by maximizing the expected complete-data log-pseudo-likelihood:

\begin{align}
\boldsymbol{\Omega}^{(s, t+1)}=\argmax _{\boldsymbol{\Omega} \in \boldsymbol \Theta} \mathbb{E}_{z_i \sim \text{Cat}(\tau_{i \ell}^{(s, t+1)})}\left\{\log f\left(\mb{A}, \boldsymbol{z} ; \boldsymbol{\Omega}^{(s, t)}, \boldsymbol{e}^{(s)}\right)\right\},
\end{align}
where  
\begin{align*}
f\left(\mb{A}, \boldsymbol{z} ; \boldsymbol{\Omega}^{(s, t)}, \boldsymbol{e}^{(s)}\right)=\prod_{i=1}^n\left\{\pi_{z_i} \prod_{j=1}^n  \left(  \left\{Q_{z_i e_j^{(s)}}^{(s, t)}\right\}^{B_{i j}}\left\{1-Q_{z_i e_j^{(s)}}^{(s, t)}\right\}^{1-B_{i j}} P_{z_i e_j^{(s)}}^{(s, t)}
  \right)^{|A_{ij}|} \left( 1 - P_{z_i e_j^{(s)}}^{(s, t)} \right)^{1 - |A_{ij}|}  \right\} .
\end{align*}
The updates of $\boldsymbol{\pi}^{(s,t+1)}$ and $\mb{P}^{(s,t+1)}$ have closed forms: for $1 \le \ell, \ell' \le K$,
\begin{equation}
    \pi_\ell^{(s, t+1)}  =\frac{1}{n} \sum_{i=1}^n \tau_{i \ell}^{(s, t+1)} \quad \text{ and}\quad P_{\ell, \ell^{\prime}}^{(s, t+1)}=\frac{T_{\ell, \ell^{\prime}}^{(s, t+1)} + S_{\ell, \ell^{\prime}}^{(s, t+1)}}{T_{\ell, \ell^{\prime}}^{(s, t+1)} + S_{\ell, \ell^{\prime}}^{(s, t+1)}+R_{\ell, \ell^{\prime}}^{(s, t+1)}},
    \label{pi}
\end{equation}
where 
\begin{align*}
& T_{\ell, \ell^{\prime}}^{(s, t+1)} = \sum_{i=1}^n \sum_{j=1}^n \tau_{i \ell}^{(s, t+1)}\cdot  I\left\{A_{i j} = 1\right\}  \cdot I\left\{e_j^{(s)}=\ell^{\prime}\right\}, \\
& S_{\ell, \ell^{\prime}}^{(s, t+1)} = \sum_{i=1}^n \sum_{j=1}^n \tau_{i \ell}^{(s, t+1)}\cdot I\left\{A_{i j} = -1\right\}  \cdot I\left\{e_j^{(s)}=\ell^{\prime}\right\}, \\
& R_{\ell, \ell^{\prime}}^{(s, t+1)} = \sum_{i=1}^n \sum_{j=1}^n \tau_{i \ell}^{(s, t+1)}\cdot I\left\{A_{i j} = 0\right\}  \cdot I\left\{e_j^{(s)}=\ell^{\prime}\right\}.
\end{align*}

Different from \textcite{Wang2021-pp}, the sign-probability matrix $\mb{Q}$ in BSBM is not a free parameter. Instead, it satisfies the constraint $\mb{Q} = (1+  \boldsymbol{\eta} \circ \boldsymbol{\nu} \boldsymbol{\nu}^\top )/2$ to achieve population-level balance. As a result, the update of $\mb{Q}^{(s,t+1)}$ has no closed form. 
We first obtain $\boldsymbol{\nu}^{(s,t+1)}$ by solving the following binary quadratic optimization problem, which is equivalent to a max-cut problem \parencite{Karp1972}: 
\begin{equation}
\boldsymbol{\nu}^{(s, t+1)} = \argmax_{\boldsymbol{\nu} \in\{\pm 1\}^K} \sum_{\ell=1}^K\sum_{\ell'=1}^K \nu_\ell \cdot \text{sign}\left(T_{\ell, \ell^{\prime}}^{(s, t+1)} - S_{\ell, \ell^{\prime}}^{(s, t+1)}\right)\cdot  U_{\ell, \ell'}^{(s, t+1)} \cdot \nu_{\ell'},
    \label{v}
\end{equation}
where $U_{\ell, \ell'}^{(s, t+1)} = T_{\ell, \ell^{\prime}}^{(s, t+1)} \log \frac{2T_{\ell, \ell^{\prime}}^{(s, t+1)}}{T_{\ell, \ell^{\prime}}^{(s, t+1)} + S_{\ell, \ell^{\prime}}^{(s, t+1)}} + S_{\ell, \ell^{\prime}}^{(s, t+1)} \log \frac{2S_{\ell, \ell^{\prime}}^{(s, t+1)} }{T_{\ell, \ell^{\prime}}^{(s, t+1)} + S_{\ell, \ell^{\prime}}^{(s, t+1)}}$.

When $K$ is small, we solve the optimization problem in (\ref{v}) exactly via exhaustive search. For larger $K$, we apply the Goemans-Williamson algorithm \parencite{goemans1995improved}, which solves a semi-definite programming relaxation of the max-cut problem to obtain an efficient approximation of the optimal solution.
Given $\boldsymbol{\nu}^{(s, t+1)}$, the update of $\eta_{\ell,\ell'}^{(s, t+1)}$ has closed form:

\begin{equation}
    \eta_{\ell,\ell'}^{(s, t+1)}=\max \left(\frac{T_{\ell, \ell^{\prime}}^{(s, t+1)} - S_{\ell, \ell^{\prime}}^{(s, t+1)}}{T_{\ell, \ell^{\prime}}^{(s, t+1)} + S_{\ell, \ell^{\prime}}^{(s, t+1)}}\cdot \nu_{\ell}^{(s,t+1)} \nu_{\ell'}^{(s,t+1)}, 0\right).
    \label{eta}
\end{equation}
Together, we update  for $1 \le \ell, \ell' \le K$
\begin{equation} \label{Q}
    Q_{\ell,\ell'}^{(s, t+1)} =  \frac{1 +\eta_{\ell,\ell'}^{(s, t+1)} \nu_{\ell}^{(s, t+1)} \nu_{\ell'}^{(s, t+1)}}{2}.
\end{equation}

\paragraph{Outer Iterations} Given the estimates $\boldsymbol{\Omega}^{(s+1)}=(\boldsymbol{\pi}^{(s+1)}, \mb{P}^{(s+1)}, \mb{Q}^{(s+1)})$ and $\tau_{il}^{(s+1)}$ after the convergence of the inner EM algorithm, we update the column labels $\boldsymbol{e}$ by maximizing the expected complete-data log-pseudo-likelihood:
$$
\argmax_{\boldsymbol{e}\in [K]^n}\mathbb{E}_{z_i \sim \text{Cat}(\tau_{i \ell}^{(s, t+1)})}\left\{\log f\left(\mb{A}, \boldsymbol{z} ; \boldsymbol{\Omega}^{(s+1)}, \boldsymbol{e}\right)\right\}.$$ 
For computational efficiency, the update is conducted separately for each node $j$ as follows:
\begin{gather}
    \begin{aligned}
e_j^{(s+1)}=\argmax _{\ell' \in [K]} \sum_{i=1}^n \sum_{\ell=1}^K \tau_{i \ell}^{(s+1)}\biggr\{ & |A_{i j}|B_{i j} \log Q_{\ell \ell'}^{(s+1)}+|A_{i j}|\left(1-B_{i j}\right) \log \left(1-Q_{\ell \ell'}^{(s+1)}\right)  \\
& + |A_{i j}| \log P_{\ell \ell'}^{(s+1)} + \left(1- |A_{i j}| \right) \log (1 - P_{\ell \ell'}^{(s+1)})\biggr\}.
  \end{aligned}
    \label{e}
\end{gather}
 While the update in \eqref{e} does not guarantee a global maximization of the expected complete-data log-pseudo-likelihood, 
the following theorem implies that each outer iteration leads to a non-negative increase in the log-pseudo-likelihood $\mc{L}_{\mathrm{PL}}\left(\boldsymbol{\Omega}, \boldsymbol{e} ; \mb{A}\right)$ in~\eqref{eq:log-pseudo-likelihood}. 
\begin{theorem}\label{thm: converge}  Given an initial labeling vector $\boldsymbol{e}^{(0)}$, the sequence $\{(\boldsymbol{\Omega}^{(s)}, \boldsymbol{e}^{(s)})\}_{s \geq 0}$ generated by Algorithm~\ref{alg:profile_pseudo_likelihood} satisfies
    $$\mc{L}_{\mathrm{PL}}\left(\boldsymbol{\Omega}^{(s+1)}, \boldsymbol{e}^{(s+1)} ; \mb{A}\right) \geq \mc{L}_{\mathrm{PL}}\left(\boldsymbol{\Omega}^{(s)}, \boldsymbol{e}^{(s)} ; \mb{A}\right) .$$
\end{theorem}
The proof of Theorem~\ref{thm: converge} is provided in Appendix. 
Since the parameter space $\boldsymbol{\Theta}$ is compact, the objective function, i.e., the log-pseudo-likelihood function, is guaranteed to converge as the number of outer iterations increases. 
Nonetheless, since the pseudo-likelihood function is not concave, Theorem~\ref{thm: converge} does not guarantee convergence to the global optimum. The output from Algorithm~\ref{alg:profile_pseudo_likelihood} may correspond to a local maximum depending on the initialization. In practice, we find that initializing $\boldsymbol{e}^{(0)}$ using spectral clustering with perturbations \parencite{amini2013pseudo} returns reliable convergence and good empirical performance.

\section{Strong Consistency}
\label{sec:theory}
 In this section, we establish the strong consistency of the estimator obtained from our proposed algorithm. Here, strong consistency means that, as the network size grows, the probability of the estimated community memberships exactly match the true memberships converges to one. 
While \textcite{Wang2021-pp} showed that maximizing profile-pseudo likelihood  achieves strong consistency under the binary SBM when using only edge connectivity information, our model and the estimation algorithm are different from theirs. Our method leverages not only edge connectivity but also edge-sign information, which enables strong consistency under weaker conditions. 
For example, if each edge occurs with uniform probability, i.e., edge connections have no community signal, it becomes impossible to detect community structures based solely on edge connectivity. In contrast, in the BSBM, strong consistency can still be achieved provided there is suitable separation between the probabilities of observing positive signs within and between communities. 
To illustrate the key insights, we will start with the case of two communities with balanced sizes and then generalize our findings to any fixed number of communities.

We first consider the BSBM with two equal-sized communities. Let $m=n / 2$ denote the number of nodes in each community. 
Let the true community labels be $z_i=1$ for $1 \le i \le  m$,  $z_i=2$ for $m+1 \le i \le n$.
Without loss of generality, assume the edge-probability matrix in~\eqref{eq: edge sbm} is 

\begin{equation*}
\mb{P}=\frac{1}{m}\begin{pmatrix}
a & b \\
b & a
\end{pmatrix},\qquad \text{with} \quad 0 \le b \le a \le m,
\label{edge_prob_matrix1}
\end{equation*}
where nodes are more likely to form edges within communities than between them. 
Suppose each community maps to one meta-group, i.e., $\boldsymbol{\nu} = (1,-1)$, and let $\boldsymbol{\eta}=  \left(\begin{array}{ll}
c & d\\
d & c
\end{array}\right)$ with $c,d\in [0,1]$. Then 
the edge-sign probability matrix $\mb{Q}$ in~\eqref{sign sbm} is 
\begin{equation*}
  \mb{Q}=\frac{1}{2}\left(1+\boldsymbol{\eta} \circ \boldsymbol{\nu}\boldsymbol{\nu}^{\top}\right) = 
\frac{1}{2}\left(\begin{array}{ll}
1+c & 1-d \\
1-d & 1+c
\end{array}\right),\qquad \text{with} \quad c,d\in [0,1].
\label{sign_prob_matrix1}
\end{equation*}
Under this specification, edges are more likely to be positive within communities and negative between communities.

For the above BSBM with two communities, we establish strong consistency of the estimated community memberships after one outer iteration under mild initialization conditions. Specifically, we assume the initial column labels $\boldsymbol{e}^{(0)}\in \mathcal{P}_{\mathcal{E}}^\gamma$ and the initial model parameters $\boldsymbol{\Omega}^{(0)} \in \mathcal{P}_{\Omega}$, where 
\begin{equation*}
\mathcal{P}_{\mathcal{E}}^{\gamma}=\left\{{\boldsymbol{e}}^{(0)} \in\{1,2\}^n: \sum_{i=1}^m I\left(e_i^{(0)}=1\right)=\gamma m, \sum_{i=m+1}^n I\left(e_i^{(0)}=2\right)=\gamma m, \ \gamma \in\left(0,\frac{1}{2}\right) \cup \left(\frac{1}{2}, 1\right)\right\},
\label{gamma}
\end{equation*} and
\begin{equation*}
\begin{aligned}
\mathcal{P}_{\Omega} = \Bigg\{ \boldsymbol{{\Omega}}^{(0)}=(\boldsymbol{\pi}^{(0)}, \mb{P}^{(0)}, \mb{Q}^{(0)}): \ {\pi}_\ell^{(0)}=\frac{1}{2}, P_{\ell \ell}^{(0)} > P_{\ell \ell'}^{(0)},\  Q_{\ell \ell}^{(0)}= Q_{\ell \ell'}^{(0)}=\frac{1}{2}, 1\le \ell \neq \ell' \le 2   \Bigg \}. 
\end{aligned}
\end{equation*}
Here, the initial column labels $\boldsymbol{e}^{(0)}$ assign an equal number of nodes to each of the two communities and agree with the true community labels on $\gamma m$ nodes per community, for some $\gamma \in (0,\frac{1}{2}) \cup (\frac{1}{2}, 1)$. We assume that $\boldsymbol{e}^{(0)}$ is slightly better than random guessing (i.e., $\gamma \neq 1/2$) and exclude the degenerate cases $\gamma = 0$ or $\gamma = 1$, which already correspond to perfect labeling up to permutation. 

The assumptions on the initial model parameters $\boldsymbol{{\Omega}}^{(0)}$ are mild. The initial within-community edge probability is greater than the between-community probability, the initial community proportions are balanced, and the sign-probability matrix is not required to be informative. 

Given the initial labels $\boldsymbol{e}^{(0)}$ and parameters~${\boldsymbol{\Omega}}^{(0)}$, one outer iteration first computes the posterior probabilities $\hat{\tau}_{i \ell}$ for $i \in [n], \ell \in \{1,2\}$ using~\eqref{tau} and then updates the column labels following~\eqref{e}. We denote the estimated labels after one outer iteration by $\hat{\boldsymbol{z}} (\boldsymbol{e}^{(0)}, \boldsymbol{\Omega}^{(0)} )$. The following theorem establishes the strong consistency of this one-step estimator. 

\begin{theorem}\label{consistencyk2}
Suppose $ \frac{(a(1+c) - b(1-d))^2}{a(1+c) + b(1-d)} \geq C \log n $, $ \frac{(a(1-c) - b(1+d))^2}{a(1-c) + b(1+d)} \geq C \log n $, and $ac+bd \geq C \log n$ hold for sufficiently large constant $C>0$, 
then for any $\epsilon > 0$, there exists $N > 0$ such that for all $n \geq N$,
$$
\begin{aligned}
 \mathbb{P} & \left\{\bigcap_{\boldsymbol{\Omega}^{(0)} \in \mathcal{P}_{\Omega}}\left\{\hat{\boldsymbol{z}}\left(\boldsymbol{e}^{(0)}, \boldsymbol{\Omega}^{(0)}\right)=\boldsymbol{z}\right\}\right\}\geq  1-  \left[n e^{-\frac{(a(1+c) - b(1-d) -4\epsilon)^2}{8(a(1+c) + b(1-d))} } + n e^{-\frac{(a(1-c) - b(1+d) -4\epsilon)^2}{8(a(1-c) + b(1+d))} } \right.\\
& \ \ \ \ \ \ \ \ \ \ \ \ \ \ \ \ \ \ \ \left.  +(2n^2 + 2n) \left( e^{ - \frac{(2 \gamma-1)^2(a(1+c)-b(1-d))^2}{16(a(1+c)+b(1-d))}   }  +  e^{ - \frac{(2 \gamma-1)^2(a(1-c)-b(1+d))^2}{16(a(1+c)+b(1-d))}   } +  e^{-  \frac{(2 \gamma-1)^2(ac +bd)}{8} }\right)\right],
\end{aligned}
$$
 for any $\boldsymbol{e}^{(0)} \in \mathcal{P}_{\mathcal{E}}^{\gamma}$, 
where $\hat{\boldsymbol{z}}=\boldsymbol{z}$ denotes equivalence up to a permutation of community labels. 
\end{theorem}
The proof of Theorem~\ref{consistencyk2} is given in Appendix. As $n \to \infty$, the bound in Theorem~\ref{consistencyk2} implies that the one-step estimator $\hat{\boldsymbol{z}}$ is a strongly consistent estimate of community labels. 
Our result for strong consistency requires $
\frac{(a(1+c) - b(1-d))^2}{a(1+c) + b(1-d)} \geq C \log n
$ and $
\frac{(a(1-c) - b(1+d))^2}{a(1-c) + b(1+d)} \geq C \log n
$. 

Note that the probabilities of forming positive edges are $\frac{a(1+c)}{n}$ within communities and $\frac{b(1-d)}{n}$ between communities, while those for negative edges are $\frac{a(1-c)}{n}$ within communities and $\frac{b(1+d)}{n}$ between communities. Therefore, our result requires a sufficient gap between the within- and between-community probabilities for both positive and negative edges, which shares the same order as the gap condition required for strong consistency in the binary SBM setting~\parencite{Wang2021-pp}. Both conditions are necessary in our method, as the pseudo-likelihood of the BSBM involves randomness from both edge formation and sign generation. In addition, we require $ac+bd \geq C \log n$. When $c =d$, this simplifies to $(a+b)c  \geq C \log n$, where $a+b$ corresponds to the average node degree (often denoted by $\lambda_n$ in the literature~\parencite{bickel2009nonparametric,zhao2012consistency}). Intuitively speaking, when the signal strength from the edge-sign distribution, reflected by the magnitude of $c$, is weak, a denser network is needed to ensure that enough signed edges are observed. Conversely, when the signal from edge signs is sufficiently strong, i.e., $c$ is of constant order, the condition reduces to requiring $\lambda_n \geq C \log n$, which aligns with the standard requirement $\lambda_n /\log n \rightarrow \infty$ for strong consistency in SBMs. Notably, in this case, we do not require any gap in the edge-formation probabilities between and within communities, even when $a=b$, strong consistency can still be achieved as long as the sign information provides sufficient signal (see further discussion in Remark~\ref{rmk: theory} below).

\begin{remark}\label{rmk: theory}
If community detection relies solely on edge connectivity patterns, strong consistency under the binary SBM with two balanced communities requires $\frac{(a-b)^2}{a+b} \geq C \log n$ for a sufficiently large constant $C$ \parencite{Wang2021-pp}. This condition excludes the degenerate case $a=b$, where within- and between-community edge probabilities are identical and thus contain no community signal. 
Our result extends this boundary by allowing $a=b$.
 
When $a = b$ and $c=d$, our conditions reduce to $ac^2 \geq C \log n$ for a sufficiently large constant $C$. In this setting, the problem is equivalent to fitting an SBM where each observed edge takes value $1$ with probability $(1+c)/2$ and $-1$ with probability $(1-c)/2$, and each edge is observed independently with probability $p = a/m$. Note that when $c$ is of constant order, \textcite{mariadassou2020consistency} shows that consistency for SBMs with missing data requires the sampling probability to satisfy $p \gg \log(n) / n$; under this regime, our condition naturally holds. Our theory generalizes these results by establishing strong consistency of the estimated memberships as long as $pc^2 \gg \log(n) / n$.
\end{remark}

Next, we generalize our results to $K$ equal-sized communities. Let $m = n/K$ denote the number of nodes in each community and the true community labels be $z_i = k$ for $i \in I_k= \{ (k-1)m+1, \ldots, km \}$.
Without loss of generality, assume the edge probability in~\eqref{eq: edge sbm} is
\begin{equation*}
 {P}_{\ell \ell'}=\frac{a}{m} I(\ell = \ell')+\frac{b}{m} I(\ell \neq \ell'), \quad \text{for} \quad  1\le  \ell,\ell'\le K,
\label{edge_prob_matrixK}
\end{equation*}
where nodes are more likely to form edges within communities than between them.  Suppose each community maps to one of two meta-groups via $\nu_{\ell} = (-1)^{\ell + 1}$ for $1 \le \ell  \le K$, and let $\eta_{\ell \ell'} = c \cdot  I(\ell = \ell') + d \cdot I(\ell \neq \ell')$ with $c,d\in [0,1]$ in~\eqref{sign sbm}. 

Then 
the edge-sign probability matrix $\mb{Q}$ takes $(1+c)/2$ within communities and $(1-d)/2$ or $(1+d)/2$ between communities.  

Similarly, we assume the initial column labels $\boldsymbol{e}^{(0)}\in \mathcal{P}_{\mathcal{E}}^\gamma$ and the initial model parameters $\boldsymbol{\Omega}^{(0)} \in \mathcal{P}_{\Omega}$, where 
\begin{align*}
\mathcal{P}_{\mathcal{E}}^{\boldsymbol{\gamma}}=\left\{{\boldsymbol{e}}^{(0)} \in[K]^n: \sum_{i \in I_k} I\left(e_i^{(0)}=k\right)=\gamma_k m, \ \sum_{i=1}^n I\left(e_i^{(0)}=k\right)=m, \ \gamma_k \in \left(\frac{1}{2}, 1\right), \ k \in [K]\right\},
\label{gamma}
\end{align*} and
\begin{equation*}
\begin{aligned}
\mathcal{P}_{\Omega} = \Bigg\{ \boldsymbol{{\Omega}}^{(0)}=(\boldsymbol{\pi}^{(0)}, \mb{P}^{(0)}, \mb{Q}^{(0)}): \ {\pi}_\ell^{(0)}=\frac{1}{K}, P_{\ell \ell}^{(0)} > P_{\ell \ell'}^{(0)},\  Q_{\ell \ell}^{(0)}= Q_{\ell \ell'}^{(0)}=\frac{1}{2}, 1\le \ell \neq \ell' \le K   \Bigg \}. 
\end{aligned}
\end{equation*} 

The following corollary establishes the strong consistency of the one-step estimator. 
\begin{corollary}\label{consistencyk3}
Suppose $S_1:=\frac{(a(1+c)-b(1-d))^2}{(a(1+c)+b(1-d))} \geq C \log n
$, $S_2:=\frac{(a(1-c)-b(1+d))^2}{(a(1-c)+b(1+d))} \geq C \log n$, $S_3:= \frac{(a(1+c)-b(1+d))^2}{(a(1+c)+b(1+d))}  \geq C \log n$, $S_4:= \frac{(a(1-c)-b(1-d))^2}{(a(1-c)+b(1-d))} \geq C \log n$, and $S_5:= ac-bd \geq  C\log n$, for sufficiently large constant $C>0$,  then for any $\epsilon > 0$, there exists $N > 0$ such that for all $n \geq N$, 
\begin{align*}
 \mathbb{P}& \left\{\bigcap_{\boldsymbol{\Omega}^{(0)} \in \mathcal{P}_{\Omega}} \left\{\hat{\boldsymbol{z}}\left(\boldsymbol{e}^{(0)},\boldsymbol{{\Omega}}^{(0)}\right)=\boldsymbol{z}\right\}\right\} \\
 & \geq 1-  n(K-1) \left[e^{\left(-\frac{(a(1+c)-b(1-d)-4\epsilon)^2}{16(a(1+c)+b(1-d))} \right)}  + e^{\left(-\frac{(a(1-c)-b(1+d)-4\epsilon)^2}{16(a(1-c)+b(1+d))}  \right)} \right.\\
& \ \ \ \ \ \ \ \ \ \ \ \ \ \ \ \ \ \ \ \left.+ e^{\left(-\frac{(a(1+c)-b(1+d)-4\epsilon)^2}{16(a(1+c)+b(1+d))}  \right)} + e^{\left(-\frac{(a(1-c)-b(1-d)-4\epsilon)^2}{16(a(1-c)+b(1-d))} \right)}    \right]  \\
& \ \ \ \ \ \  \ \ \ \ \ \ \   -  \frac{(16K-8)n^2}{K}  \sum_{\ell=1}^K\sum_{\ell'=1}^K \sum_{w = 1}^5 \exp \left(  - \frac{S_w}{24} (r_{\ell} + r_{\ell'} - 1)\right)  ,
\end{align*}
for any $\boldsymbol{e}^{(0)} \in \mathcal{P}_{\mathcal{E}}^{\gamma}$,
where $\hat{\boldsymbol{z}}=\boldsymbol{z}$ denotes equivalence up to a permutation of community labels.
\end{corollary} 

The proof of Corollary~\ref{consistencyk3} is provided in Appendix, which is a natural extension of the proof of Theorem~\ref{consistencyk2}.  
Compared with Theorem \ref{consistencyk2}, we require additional conditions $\frac{(a(1+c) - b(1+d))^2}{a(1+c) + b(1+d)} \geq C \log n$ and $\frac{(a(1-c) - b(1-d))^2}{a(1-c) + b(1-d)} \geq C \log n$.
These additional conditions arise from the presence of mixed entries in the between-community sign-probability matrix when $K > 2$.
Specifically, when $K>2$, the off-diagonal entries of the sign-probability matrix $Q$ can take either $(1-d)/2$ or $(1+d)/2$, while the diagonal entries are $(1+c)/2$. As a result, the probabilities of forming positive edges are $\frac{a(1+c)}{n}$ within communities, and $\frac{b(1-d)}{n}$ or $\frac{b(1+d)}{n}$ between communities, while those for negative edges are $\frac{a(1-c)}{n}$ within communities, and $\frac{b(1+d)}{n}$ or $\frac{b(1-d)}{n}$ between communities. This difference among between-community probabilities necessitates that the minimum separation between within- and between-community probabilities  be sufficiently large.
Furthermore, when $a=b=\lambda_n$, the condition $ac - bd\geq  C \log n $ simplifies to $\lambda_n(c-d)\geq C \log n$. In addition to requiring the average node degree $\lambda \gg \log n$, we also need $c>d$ strictly, i.e., the probabilities of forming positive edges are strictly higher within communities than between them, since the latter may take value $(1+d)/2$.

\section{Simulation Studies}
\label{sec:simulation}
\subsection{Simulation Setup}
We compare our method with four baselines for community detection in signed networks,  which differ in how they use edge signs and connectivity patterns. 
\begin{itemize}
\item The \textbf{low-rank matrix completion (MC)} method relies solely on \emph{edge-sign} information and does not use \emph{edge-connectivity} patterns for community detection. It treats zeros in the signed adjacency matrix as missing values and completes the matrix under a low-rank assumption using the observed edge signs~\parencite{chiang2014prediction}. Community assignments are obtained via spectral embedding followed by $k$-means clustering. 

\item The \textbf{spectral clustering with perturbation (SCP)}~\parencite{amini2013pseudo} and \textbf{profile-pseudo likelihood (PPL)} \parencite{Wang2021-pp} methods for binary  networks detect communities based solely on  \emph{edge-connectivity} and ignore sign patterns. We binarize the signed network by retaining only edge connectivity information, i.e., treating $|\mb{A}|$ as a binary unsigned adjacency matrix. The former applies the SCP algorithm on the binary network, while the latter fits an SBM for binary network via PPL estimation.

\item The \textbf{PPL-merge} baseline partially preserves sign information by merging negative edges with non-edges, i.e., setting \(-1\) entries to \(0\) in the adjacency matrix, and then fitting an SBM for the resulting binary network via profile pseudo-likelihood estimation. This baseline is motivated by the assumption that negative edges tend to occur across communities.
\end{itemize}

We generate signed networks from BSBM. The community membership of each node $z_i \in [K]$ is drawn i.i.d.\ from a categorical distribution with probabilities $\boldsymbol{\pi}=(\pi_1, \cdots, \pi_K)$. We set balanced communities $\pi_k = 1/K$ for $k \in [K]$ in the main text and provide additional results for unbalanced communities in Appendix ~\ref{app:simulation}. 
Given community memberships, edges are formed independently with probabilities $P_{z_i  z_j}$, where 
$P_{\ell \ell'}= P_{bt} + (P_{in} - P_{bt}) \cdot I(\ell=\ell')$ for $1\le \ell, \ell' \le K$ with $P_{in}=0.13$ and $P_{bt}=0.07$ by default. 
Conditional on an edge between nodes $i$ and $j$, its sign is positive with probability $Q_{z_i  z_j}$, where $Q_{\ell \ell'} = \left( 1 + \eta_{\ell \ell'}\cdot  \nu_{\ell} \cdot \nu_{\ell'}\right)/2$ for $1\le \ell, \ell' \le K$ with $\eta_{\ell \ell'}$  randomly generated from the uniform distribution $U([0,1])$, and $\nu_{\ell} = (-1)^{\ell+1}$ by default.  
We set the number of nodes $n=1,000$ and the number of communities $K = 3$ unless otherwise stated. 

To evaluate community detection performance, we report normalized mutual information (NMI) between estimated and true community memberships. It ranges from 0 (independent assignment) to 1 (perfect match); a higher value indicates a better community detection accuracy. We report results for $100$ independently generated signed networks. 

In the following subsections, we investigate four settings that vary (1) the signal from edge-connectivity patterns, quantified by the gap between within- and between-community connection probabilities $P_{in} - P_{bt}$;
(2) the meta-group size, i.e., the number of communities mapped to the same meta-group;
(3) the signal from edge-sign patterns, quantified by the magnitude of $\eta_{\ell \ell'}$ in $\mb{Q}$; 
 and
(4) the network size $n$ and the number of communities $K$.

\subsection{Varying the Signal from Edge-Connectivity Patterns}
In this setting, we investigate how the strength and direction of the edge-connectivity signal, quantified by the gap between within- and between-community connection probabilities, affect the performance of the five methods. The between-community probability is fixed at $P_{\text{bt}}=0.07$, while the within-community probability $P_{\text{in}}$ varies from $0.05$ to $0.13$. 

\begin{figure}[!htbp]
\centering
\includegraphics[width=0.9\textwidth]{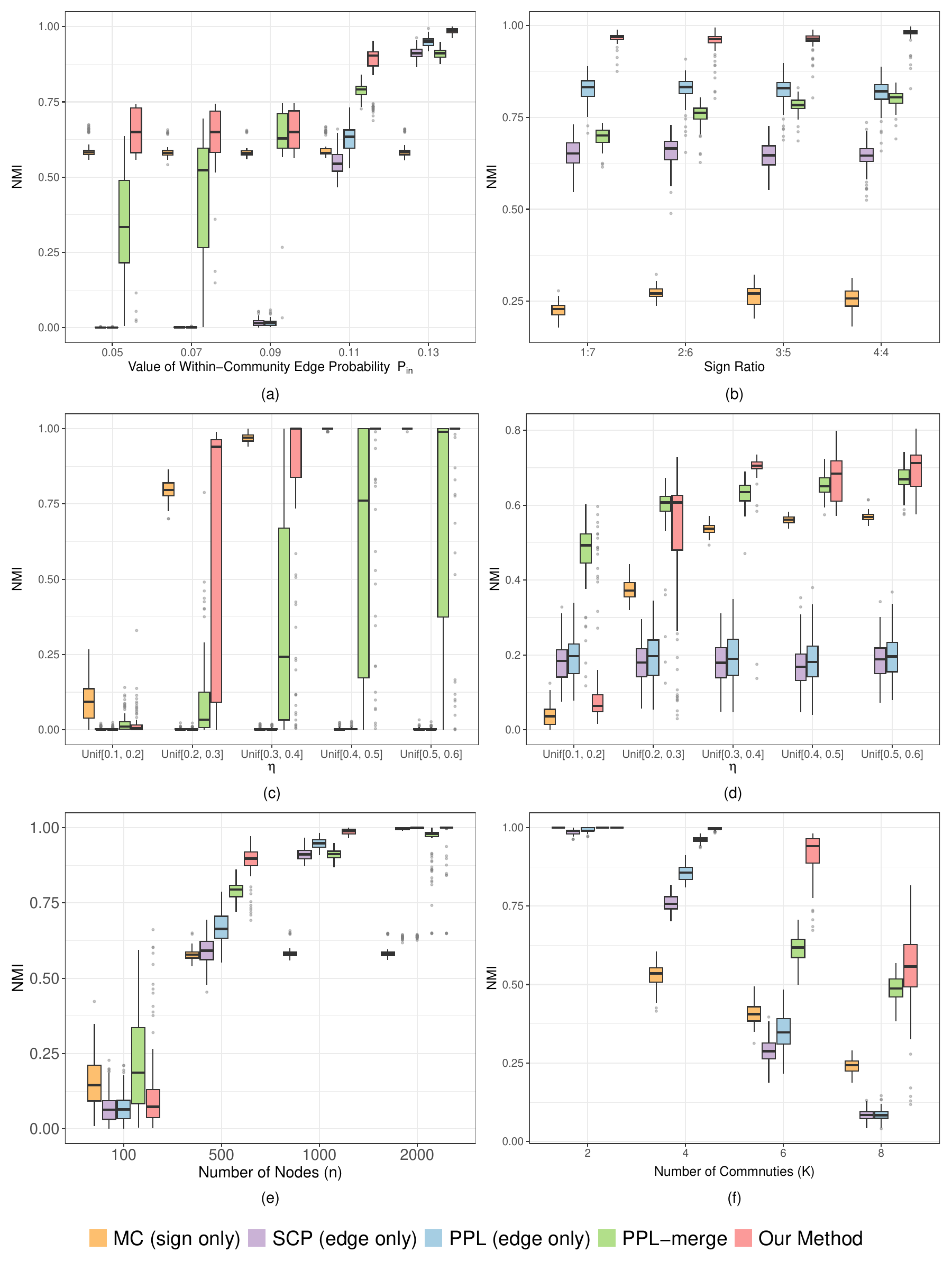}
\caption{Comparisons of NMI scores across five methods under six simulation settings. Panels (a)-(f), from top left to bottom right, correspond to: (a) varying within-community probabilities $P_{in}$ with fixed between-community probabilities $P_{bt} = 0.07$; (b) varying the size of meta-groups; (c) varying the magnitude of $\eta$'s when $K = 2$; (d) varying the magnitude of $\eta$'s when $K = 3$; (e) varying the number of nodes $n$; and (f) varying the number of communities $K$. 
}
\label{fig:simulation}
\end{figure}

Figure~\ref{fig:simulation}(a) shows that, as $P_{\text{in}}$ increases, the NMI scores of all methods except MC improve, which demonstrates improved detectability with a larger gap between within- and between-community probabilities. 
The performance of the MC method remains flat, as it relies solely on edge signs and ignores connectivity information. Across all values of $P_{\text{in}}$, our proposed method consistently outperforms the others. Notably, when $P_{\text{in}}$ is close to $P_{\text{bt}}$, corresponding to an extremely weak edge-connectivity signal, the NMI scores of PPL and SCP drop to nearly zero, indicating their failure to recover meaningful communities when relying only on connectivity information. In contrast, our method, MC, and PPL-merge achieve substantially higher NMI scores in this challenging scenario by effectively leveraging sign information. 

In addition, when $P_{\text{in}}$ further decreases into the disassortative regime ($P_{\text{in}}=0.05 < P_{\text{bt}}$), the performance of PPL-merge drops. This is because, in disassortative networks, negative edges tend to occur between communities, while non-edges are more likely to appear within communities. Merging these two types of edges therefore conflates distinct patterns. 
These results demonstrate the importance of appropriately incorporating edge-sign patterns, especially when the signal from edge-connectivity patterns is weak. 

\subsection{Varying the Meta-Group Size}

We further investigate the performance of the five methods when varying the \emph{sign ratio} in the vector $\nu$, which determines how many communities share the same sign and thus belong to the same meta-group. We set the number of communities $K = 8$ with equal community size. 
A sign ratio of 1:7, for example,  corresponds to the case where only the first entry of $\nu$ is $-1$, and all remaining seven entries are $1$, indicating one meta-group is seven times larger than the other. A more balanced sign ratio 4:4 implies that communities are evenly divided between the two meta-groups. 

The within- and between-community probabilities $P_{in}$ and $P_{bt}$ are fixed at $0.15$ and $0.06$, respectively. We choose a relatively high $P_{\text{in}}$ and a larger gap between $P_{\text{in}}$ and $P_{\text{bt}}$ than the default values since the case with $K=8$ is more challenging. As shown in Figure~\ref{fig:simulation}(b), our method remains robust and consistently outperforms the other methods across different sign ratios.

\subsection{Varying the Signal from Edge-Sign Patterns}
In this setting, we evaluate the community detection performance of the five methods under varying strengths of the edge-sign signal. The entries of the edge-sign probability matrix $Q$ are randomly generated from uniform distributions with different ranges to control the signal magnitude. Specifically, $\eta_{\ell, \ell'}$ are uniformly sampled from five different intervals: $U([0.1, 0.2])$, $U([0.2, 0.3])$, $U([0.3, 0.4])$, $U([0.4, 0.5])$, and $U([0.5, 0.6])$, respectively. Larger values correspond to stronger signals from edge-sign patterns.
We set $P_{\text{bt}} = 0.07$ and $P_{\text{in}} = 0.1$, slightly lower than the default value, to create a more challenging scenario where differences in performance across methods become clearer.

Figures~\ref{fig:simulation}(c) and~\ref{fig:simulation}(d) present the results for $K=2$ and $K=3$, respectively. In both cases, the performances of our method, MC, and PPL-merge improve as the magnitude of $\eta$'s increases, which benefits from stronger signals from edge signs. The performances of SCP and PPL, which only use edge-connectivity information, remain almost identical. When $K = 2$, the structure of~$\mb{Q}$ ensures that the edges within community are more likely to be positive and the edges between communities between are more likely to be negative. In this scenario, methods that leverage sign information significantly outperform those that do not. For $K = 3$, when the values of $\eta$'s are small, the probabilities of edge signs between and within communities become less distinguishable, which reduces the informativeness of signs patterns. Consequently, the performance of our method decreases under weak sign signals. However, as shown in Figure~\ref{fig:simulation}(d), when the values exceed 0.3, our method achieves the best performance among all five methods.

\subsection{Varying the Network Size $n$ and the Number of Communities $K$}
Finally, we study the impact of varying the network size and the number of communities. We first fix $K=3$ and vary the number of nodes from $100$ to $2,000$. As shown in Figure~\ref{fig:simulation}(e), the performances of all methods improve as $n$ increases. Our method consistently outperforms all baselines when $n \geq 500$; and when $n = 2,000$, our method, SCP, PPL, and PPL-Merge all achieve near-perfect community recovery. For small networks with $n=100$, the performance of our method declines because it is subject to two sources of randomness arising from edge formation and sign generation, whereas the baselines rely on only one. The additional randomness in our method makes community recovery more difficult given the limited network size.

Next, we fix the network size $n=1,000$ and vary the number of communities $K$ over $\{2,4,6,8\}$. Figure~\ref{fig:simulation}(f) shows that NMI generally decreases as $K$ increases, which is expected as the community detection problem becomes more challenging. Nonetheless, our proposed method consistently outperforms all baselines across different values of $K$. When $K = 2$, all five methods achieve NMI scores close to $1$. As $K$ increases, the performances of SCP and PPL deteriorate substantially, with NMI scores falling below $0.2$ when $K = 8$. MC performs slightly better but still lags behind our method and PPL-merge.

\section{Real-world Data Examples}
\label{sec:realdata}
\subsection{International Relation Network}
We apply our method to the \textit{Correlates of War} (COW) dataset, which records  various types of international relations between countries, including wars, alliances, and militarized interstate disputes~\parencite{izmirlioglu2017correlates}. Following the procedures in~\textcite{tang2025population}, we construct a signed international relations network by treating alliance records as positive relationship and records of wars or militarized disputes as negative relationship. We focus on the period from 1941 to 1943, corresponding to the phase of World War II after the U.S. entered the conflict, during which the major powers largely maintained stable alliance structures. For each pair of countries, we aggregate records over this three-year period. The sign of the edge is set to negative if the duration of negative relationship is longer than that of positive relationship and set to positive otherwise. Country pairs with no recorded interactions during this period are assigned non-edge. The signed network includes 52 countries, with 237 positive edges and 184 negative edges in total. 
We assess the empirical evidence of structural balance theory in this signed network using the hypothesis test proposed by \textcite{feng2022testing}. The p-value is near 0, which provides strong evidence against the null hypothesis of a balance-free network, and motivates us to leverage balance theory for community detection.  
We adopt the network cross-validation method \parencite{li2020network} and select the number of communities $K=8$.

\begin{figure}[H]
\centering
\includegraphics[width=0.7\textwidth]{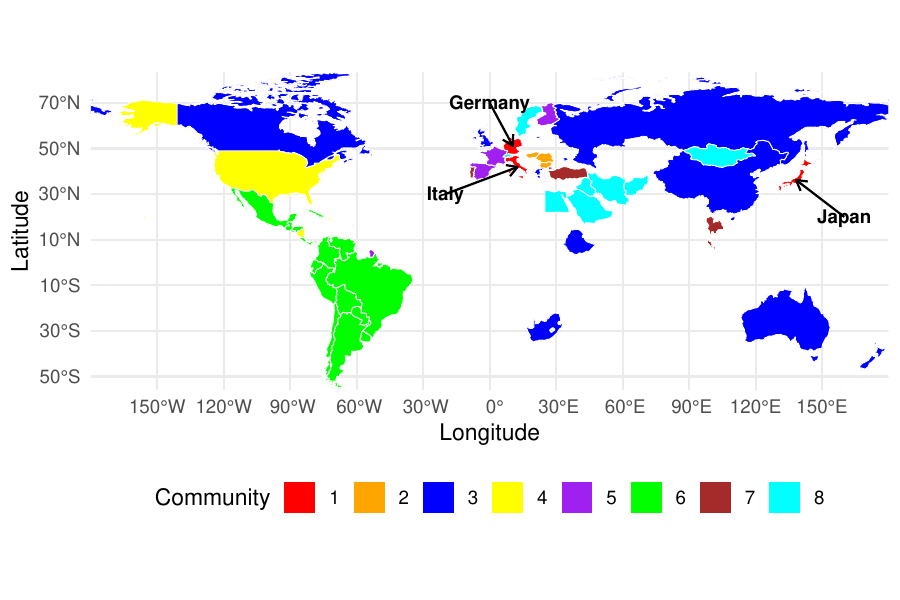}
\caption{World map showing eight communities identified by our method, with each color representing one community: (1) Germany, Italy, Japan;
(2) Hungary, Bulgaria, Romania;
(3) Canada, United Kingdom, Yugoslavia, Greece, Russia, Ethiopia, South Africa, China, Australia, New Zealand;
(4) United States, Haiti, Nicaragua;
(5) France, Spain, Finland;
(6) Cuba, Dominican Republic, Mexico, Guatemala, Honduras, El Salvador, Costa Rica, Panama, Colombia, Venezuela, Ecuador, Peru, Brazil, Bolivia, Paraguay, Chile, Argentina, Uruguay;
(7) Portugal, Turkey, Thailand;
(8) Switzerland, Sweden, Iran, Iraq, Egypt, Saudi Arabia, Yemen Arab Republic, Afghanistan, Mongolia. Different colors represent different communities. 
}
\label{fig:world map}
\end{figure}

Our method identifies eight communities that align closely with geopolitical structures during 1941–1943 (see Figure~\ref{fig:world map} for detailed assignments).
Community~1 captures the core Axis powers, and Community~2 includes their Eastern European allies.  
The Allied powers are divided into multiple communities: one centered on the United Kingdom and Commonwealth nations (Community~3), another on the United States and its regional allies (Community~4), and a third including France, Spain, and Finland (Community~5).

Latin American countries cluster together in Community~6, reflecting regional ties rather than direct military engagement. 
Neutral or less involved states, such as Switzerland, Sweden, and several Middle Eastern countries, form Community~8, partially due to their geopolitical distance from the primary conflict. Overall, our method effectively captures meaningful political and geographical community structures.

For comparison, we also apply PPL and PPL-merge to the same dataset. Both methods partially recover geopolitical patterns but produce less coherent clusters than ours. Detailed community assignments for both methods are reported in Appendix~\ref{app:real data}. 
Under PPL, Hungary, Bulgaria, and Romania are grouped with the Commonwealth nations, which is inconsistent with the fact that the former were Axis-aligned while the latter were Allied supporters. In addition, the United Kingdom is grouped with Portugal, Iran, Turkey, and Thailand, which appears inconsistent with historical records, as both Portugal and Turkey maintained neutrality most of the time. Under PPL-merge, Canada, France, China, and the Commonwealth nations are grouped with Italy, which was one of major Axis powers, while Germany is grouped together with Greece, Yugoslavia, Iran, and Turkey, which were either neutral or opposed to Axis expansion. 

\subsection{Protein-Protein Interaction Network}
We further apply our method to a protein-protein interaction (PPI) network \parencite{huttlin2021dual}. In this network, each protein is represented as a node, and significant positive or negative correlations between protein expression levels define the existence of edges and their signs. We use a significance threshold of 0.05 for the correlation $p$-values and remove nodes with degrees less than 5, resulting in a network with $1,109$ nodes, $5,163$ positive edges, and $1,308$ negative edges. Similarly, we apply the hypothesis test proposed by \textcite{feng2022testing} and find strong empirical evidence of balance theory in this signed network ($p$-value $\approx 0$). The number of communities is selected as $25$ using the network cross-validation method \parencite{li2020network}.

To interpret our community detection results, we conduct an over-representation analyses (ORA) \parencite{huang2009bioinformatics} to identify biological pathways enriched within each detected community. 
The predefined sets of proteins are obtained from the Molecular Signatures Database (MSigDB)~\parencite{liberzon2015molecular}. 
Within each community, we use the hypergeometric test to assess whether predefined sets of proteins are statistically over-represented compared to what would be expected by chance. 
Multiple testing correction is applied via the Benjamini–Hochberg procedure to control the false discovery rate. 
For each community, we report the pathway with the smallest adjusted p-value. The results for our method are summarized in Table~\ref{Table:ORA result} in Appendix~\ref{app:real data}, where 23 out of 25 communities are significantly associated with biologically meaningful pathways (adjusted p-values $<0.05$).

\begin{figure}[!htbp]
    \centering
    \includegraphics[width=0.5\linewidth]{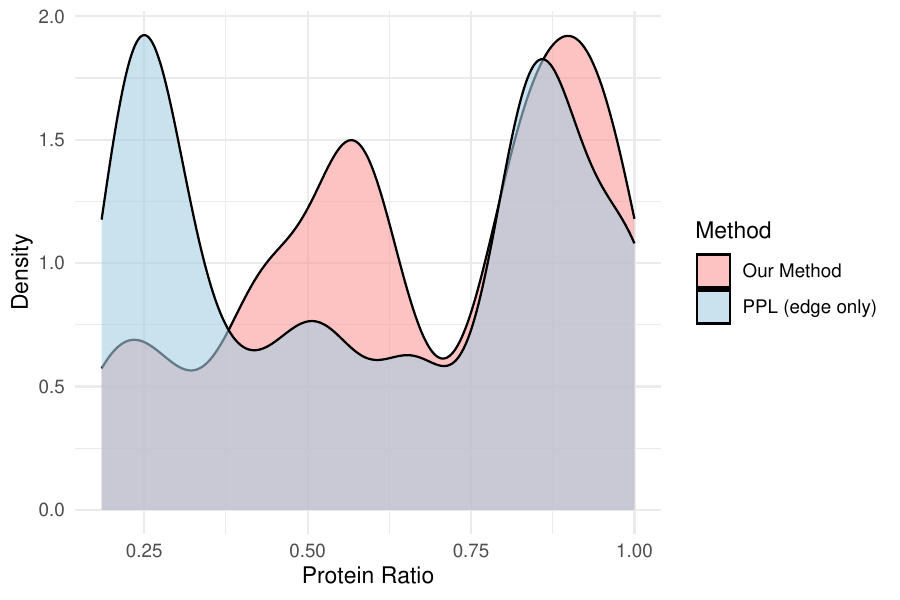}
    \caption{Density plot comparing the distribution of protein ratios obtained from our method and the PPL method. The \textit{protein ratio} ($M_1/M_2$) for a given community represents the fraction of proteins ($M_1$) annotated to the enriched pathway out of the total number of proteins in that community ($M_2$). Higher ratios indicate more coherent and functionally homogeneous communities.}
    \label{fig:protein ratio}
\end{figure}

\begin{figure}[!htbp]
    \centering
    \includegraphics[width=\linewidth]{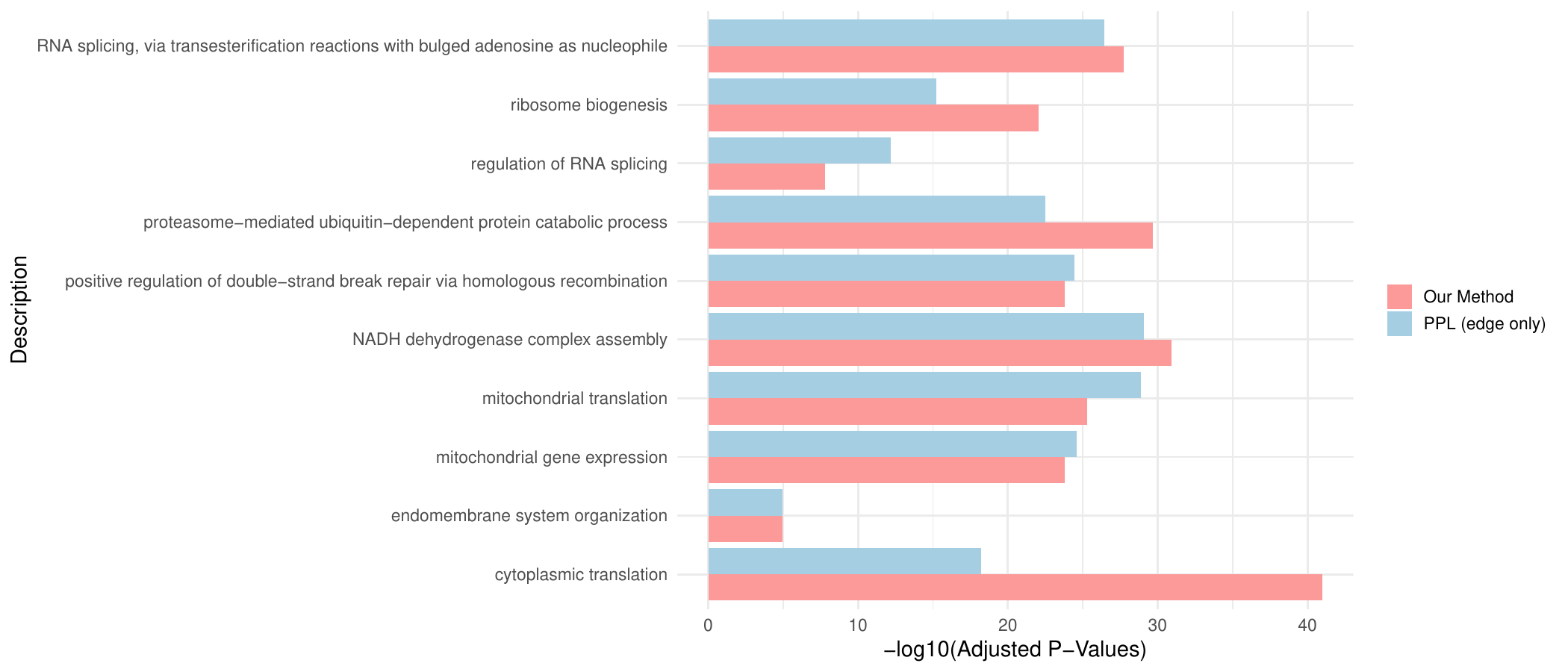}
    \caption{Comparison of over-representation analysis results between our method and the PPL method. Each bar represents $-\log_{10}$(adjusted $p$-value) for enriched biological pathway in a given community. Higher values indicate stronger statistical significance of enrichment.}
    \label{fig:bio pathway}
\end{figure}

Compared with the PPL method, which uses only edge-connectivity patterns, our method identifies communities with stronger biological interpretability. 
Specifically, Figure~\ref{fig:protein ratio} compares the distributions of \textit{protein ratios} obtained from the two methods. Here, the \textit{protein ratio} ($M_1/M_2$) for a given community represents the fraction of proteins ($M_1$) annotated to the enriched pathway out of the total number of proteins in that community ($M_2$). Higher ratios indicate more coherent and functionally homogeneous communities. As shown in Figure~\ref{fig:protein ratio}, the communities identified by our method tend to have larger protein ratios. Furthermore, for the $10$ communities with protein ratios greater than 0.5 that share the same associated pathways under both methods, Figure~\ref{fig:bio pathway} compares their negative log-transformed adjusted p-values.  Our method achieves higher values in most communities, which indicates that it identifies pathways with stronger statistical significance than those obtained using the PPL method. 
Finally, the communities obtained from the PPL method contain a higher proportion of negative edges within communities (10\% vs.\ 7\%) and a lower proportion of negative edges between communities (26\% vs.\ 28\%), indicating weaker alignment with balance theory compared to our method.

\section{Conclusion}
\label{sec:conclusion}
We have proposed a novel balanced stochastic block model (BSBM) for signed networks, which introduces a hierarchical organization of communities into two meta-groups to achieve population-level balance. 
The BSBM integrates balance theory into its generative process and leverages both edge-connectivity and edge-sign information for community detection. This design allows BSBM to uncover meaningful community structures when edge connectivity alone provides weak signals, which has been demonstrated both theoretically and numerically. 
We have also developed a fast maximum profile-pseudo likelihood estimation algorithm for fitting BSBM with convergence guarantee.  

Future work may extend our BSBM to capture diverse heterogeneity in real-world signed networks. One natural direction is to incorporate node degree heterogeneity and overlapping community memberships by adapting ideas from the degree-corrected mixed membership SBM. It is also of interest to extend BSBM to dynamic signed networks, where community memberships may evolve over time.

\clearpage
\begingroup
\setstretch{1.7}
\AtNextBibliography{\small}
\printbibliography
\endgroup

\newpage
\appendix





\section{Additional Simulation Results}
\label{app:simulation}
In this Appendix, we provide additional simulation results for unbalanced communities. The simulation set-up is consistent with the main text, except for the choice of $\boldsymbol{\pi}$.
\subsection{Varying the Signal from Edge-Sign patterns}

In this setting, we evaluate the community detection performance of the five methods under varying strengths of the edge-sign signal. The entries of the edge-sign probability matrix $Q$ are randomly generated from uniform distributions with different ranges to control the signal magnitude. Specifically, $\eta_{\ell, \ell'}$ are uniformly sampled from five different intervals: $U([0.1, 0.2])$, $U([0.2, 0.3])$, $U([0.3, 0.4])$, $U([0.4, 0.5])$, and $U([0.5, 0.6])$, respectively. Larger values correspond to stronger signals from edge-sign patterns. We set $P_{\text{bt}} = 0.07$ and $P_{\text{in}} = 0.1$, slightly lower than the default value, to create a more challenging scenario where differences in performance across methods become clearer.

\begin{figure}[htbp]
\centering
\includegraphics[width=0.8\textwidth]{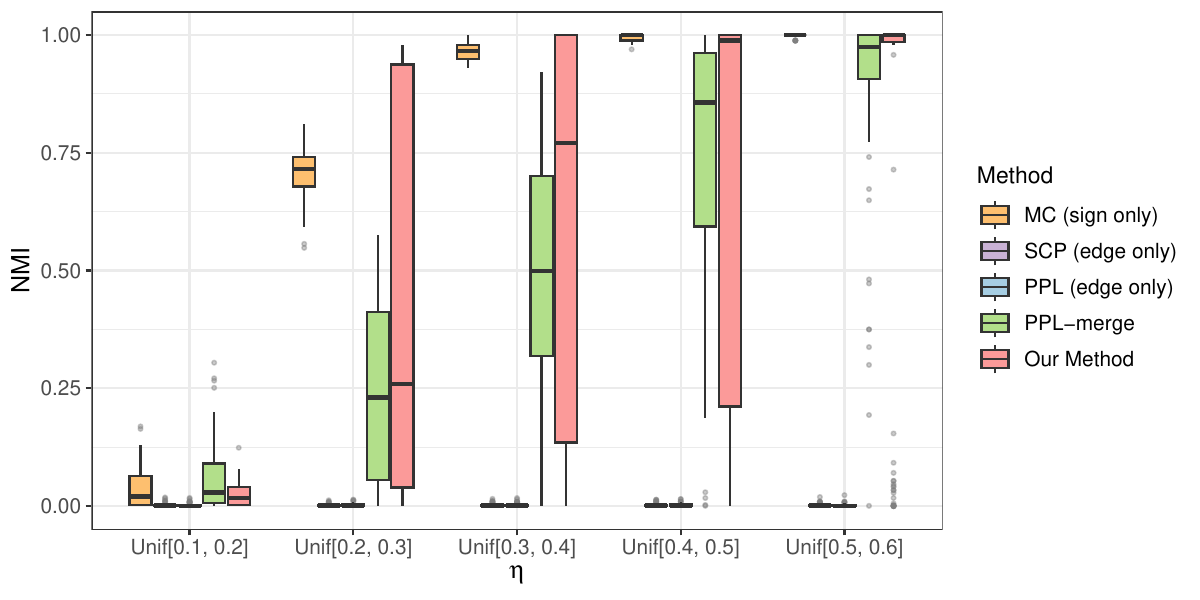}
\caption{Comparisons of NMI scores across five methods under varying magnitude of $\eta$'s when $K = 2$ with $\boldsymbol{\pi} = (0.3, 0.7). $} 
\label{fig:special eta K is 2-unbalance}
\end{figure}

\begin{figure}[htbp]
\centering
\includegraphics[width=0.8\textwidth]{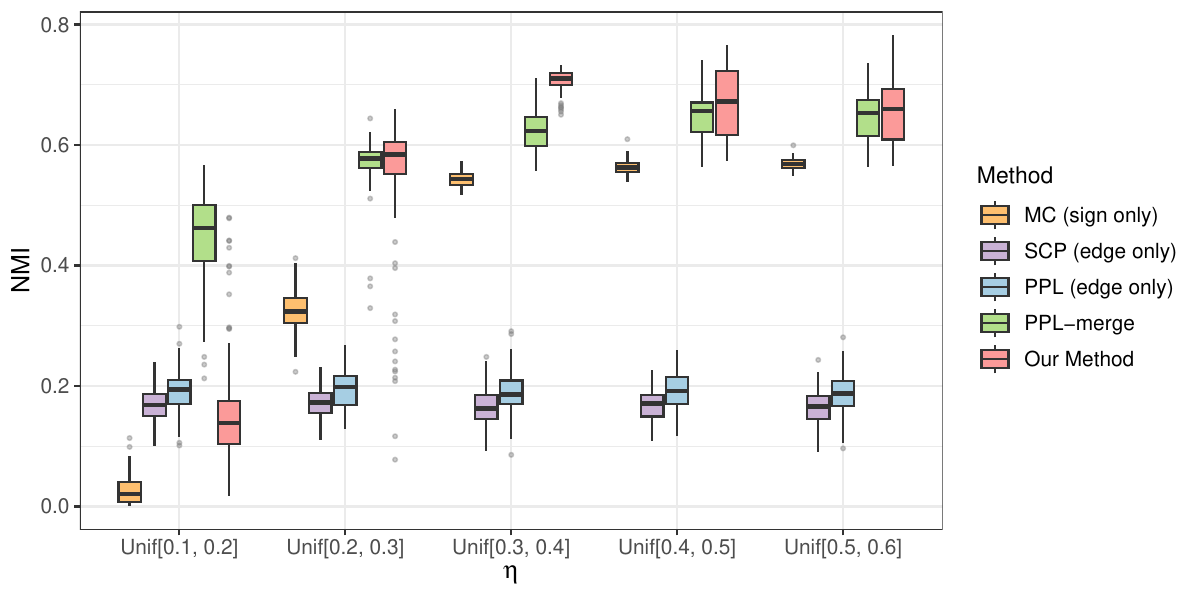}
\caption{Comparisons of NMI scores across five methods under varying magnitude of $\eta$'s when $K = 3$ with $\boldsymbol{\pi} = (0.2, 0.3, 0.5). $}
\label{fig:special eta K is 3-unbalance}
\end{figure}

Figures~\ref{fig:special eta K is 2-unbalance} and~\ref{fig:special eta K is 3-unbalance} present the results for $K=2$ and $K=3$, respectively. When $K=2$, we set $\boldsymbol{\pi} = (0.3, 0.7)$. When $K=3$, we set $\boldsymbol{\pi} = (0.2, 0.3, 0.5)$. 
In both cases, the performances of our method, MC, and PPL-merge improve as the magnitude of $\eta$'s increases, which benefits from stronger signals from edge signs. The performances of SCP and PPL, which only use edge-connectivity information, remain almost identical. When $K = 2$, the structure of~$\mb{Q}$ ensures that the edges within the community are more likely to be positive and the edges between the community are more likely to be negative. In this scenario, methods that leverage sign information significantly outperform those that do not. For $K = 3$, when the values of $\eta$'s are small, the probabilities of edge signs between and within communities become less distinguishable, which reduces the informativeness of signs patterns. Consequently, the performance of our method decreases under weak sign signals. However, as shown in Figure~\ref{fig:special eta K is 3-unbalance}, when the values exceed 0.3, our method achieves the best performance among all five methods.

\subsection{Varying the Network Size $n$ and the Number of Communities $K$}

\begin{figure}[!htbp]
\centering
\includegraphics[width=0.8\textwidth]{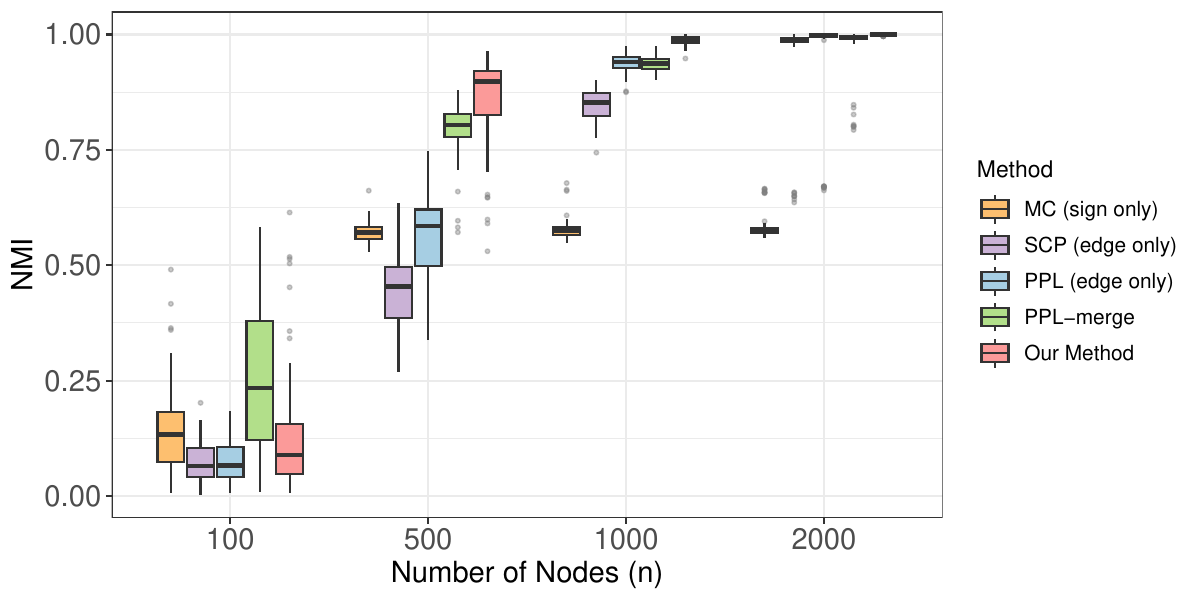}
\caption{Comparisons of NMI scores across five methods under varying number of nodes $n$ with $\boldsymbol{\pi} = (0.2, 0.3, 0.5).$}
\label{fig:varying nodes-b-unbalance}
\end{figure}

In this setting, we study the impact of varying the network size and the number of communities. We set $\boldsymbol{\pi} = (0.2, 0.3, 0.5)$. We first fix $K=3$ and vary the number of nodes from $100$ to $2,000$. As shown in Figure \ref{fig:varying nodes-b-unbalance}, the performances of all methods improve as $n$ increases. Our method consistently outperforms all baselines when $n \geq 500$; and when $n = 2,000$, our method, SCP, PPL, and PPL-Merge all achieve near-perfect community recovery. For small networks with $n=100$, the performance of our method declines because it is subject to two sources of randomness arising from edge formation and sign generation, whereas the baselines rely on only one. The additional randomness in our method makes community recovery more difficult given the limited network size.

\begin{figure}[!htbp]
\centering
        \includegraphics[width=0.8\textwidth]{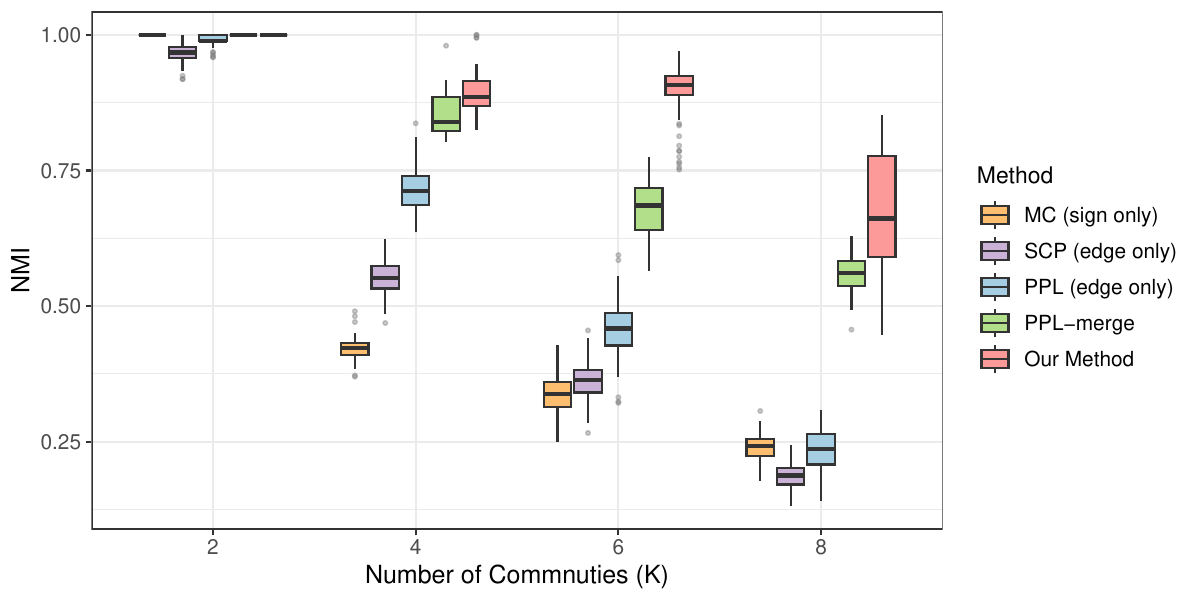}
\caption{Comparisons of NMI scores across five methods under varying number of communities $K$ with $\boldsymbol{\pi} = (0.3, 0.7)$, $\boldsymbol{\pi} = (0.1, 0.2, 0.3, 0.4)$, $\boldsymbol{\pi} =(0.05, 0.15, 0.15, 0.2, 0.2, 0.25)$, $\boldsymbol{\pi} =(0.05,0.05, 0.1,0.1,0.1,0.15,0.2,0.25)$ respectively.}.
\label{fig:varying K-a-unbalance}
\end{figure}

Next, we fix the network size $n=1,000$ and vary the number of communities $K$ over $\{2,4,6,8\}$.  We set $\boldsymbol{\pi}$ as    $\boldsymbol{\pi} = (0.3, 0.7)$, $\boldsymbol{\pi} = (0.1, 0.2, 0.3, 0.4)$, $\boldsymbol{\pi} =(0.05, 0.15, 0.15, 0.2, 0.2, 0.25)$, $\boldsymbol{\pi} =(0.05,0.05, 0.1,0.1,0.1,0.15,0.2,0.25)$, respectively. Figure~\ref{fig:varying K-a-unbalance} shows that NMI generally decreases as $K$ increases, which is expected as the community detection problem becomes more challenging. Nonetheless, our proposed method consistently outperforms all baselines across different values of $K$. When $K = 2$, all five methods achieve NMI scores close to $1$. As $K$ increases, the performances of SCP and PPL deteriorate substantially, with NMI scores falling below $0.2$ when $K = 8$. MC performs slightly better but still lags behind our method and PPL-merge.

\clearpage
\section{Additional Real Data Analyses Results}
\label{app:real data}
\subsection{International Relation Network}
The community detection results for international relation network using methods of PPL and PPL merge are shown in Table \ref{table:real data1} and Table \ref{table:real data2}.

\begin{table}[h!]
\centering
\renewcommand{\arraystretch}{1.4} 
\begin{tabular}{>{\centering\arraybackslash}m{0.8cm} >{\RaggedRight\arraybackslash}m{13cm}}
\toprule
\textbf{No.} & \textbf{Community} \\
\midrule
1 & Germany, Italy, Japan \\
2 & Yugoslavia, Greece, Russia, China \\
3 & Canada, Hungary, Bulgaria, Romania, Ethiopia, South Africa, Australia, New Zealand \\
4 & United States, Cuba, Haiti, Dominican Republic, Mexico, Guatemala, Honduras, El Salvador, Nicaragua, Costa Rica, Panama, Bolivia \\
5 & Colombia, Venezuela, Ecuador, Peru, Brazil, Paraguay, Chile, Argentina, Uruguay \\
6 & France, Spain, Finland \\
7 & United Kingdom, Portugal, Iran, Turkey, Thailand \\
8 & Switzerland, Sweden, Iraq, Egypt, Saudi Arabia, Yemen Arab Republic, Afghanistan, Mongolia \\
\bottomrule
\end{tabular}
\caption{Eight Communities identified by PPL method}
\label{table:real data1}
\end{table}

\begin{table}[h!]
\centering
\renewcommand{\arraystretch}{1.4}
\begin{tabular}{>{\centering\arraybackslash}m{0.8cm} >{\RaggedRight\arraybackslash}m{13cm}}
\toprule
\textbf{No.} & \textbf{Community} \\
\midrule
1 & Germany, Hungary, Yugoslavia, Greece, Bulgaria, Romania, Iran, Turkey, Afghanistan, Japan \\
2 & Russia, Mongolia \\
3 & Canada, France, Switzerland, Italy, Finland, Sweden, Ethiopia, South Africa, Egypt, China, Thailand, Australia, New Zealand \\
4 & (Empty) \\
5 & United States, Cuba, Haiti, Dominican Republic, Mexico, Guatemala, Honduras, El Salvador, Nicaragua, Costa Rica, Panama, Colombia, Venezuela, Ecuador, Peru, Brazil, Bolivia, Paraguay, Chile, Argentina, Uruguay \\
6 & (Empty) \\
7 & United Kingdom, Spain, Portugal \\
8 & Iraq, Saudi Arabia, Yemen Arab Republic \\
\bottomrule
\end{tabular}
\caption{Eight Communities identified by PPL-merge method}
\label{table:real data2}
\end{table}

\subsection{Protein-Protein Interaction Network}
The detailed results of over-representation analyses for our method and PPL method are presented in Table \ref{Table:ORA result} and \ref{Table:ORA result-PPL}. The \textit{protein ratio} ($M_1/M_2$) for a given community represents the fraction of proteins ($M_1$) annotated to the enriched pathway out of the total number of proteins in that community ($M_2$).  The \textit{BgRatio ($M_3/M_4$)} reflects the background distribution, where $M_3$ out of $M_4$ total proteins in the entire database are annotated with that pathway. The results of the following 10 communities are shown in Figure \ref{fig:protein ratio} in our main text: (1)RNA splicing, via transesterification reactions with bulged adenosine as nucleophile; (2)ribosome biogenesis; (3)regulation of RNA splicing; (4)proteasome-mediated ubiquitin-dependent protein catabolic process; (5)positive regulation of double-strand break repair via homologous recombination; (6)NADH dehydrogenase complex assembly; (7)mitochondrial translation; (8)mitochondrial gene expression; (9)endomembrane system organization; (10)cytoplasmic translation.

\begin{table}[!htbp]
\footnotesize
\centering
\begin{tabular}{lp{7cm}ccc}
  \toprule
Cluster & Description & ProteinRatio & BgRatio & Adjusted P-Value \\ 
  \midrule
1 & regulation of mRNA splicing, via spliceosome & 4/9 & 13/1012 & 2.80e-04 \\ 
   \midrule
2 & translational initiation & 4/9 & 17/1012 & 7.47e-04 \\ 
   \midrule
3 & positive regulation of DNA biosynthetic process & 8/35 & 11/1012 & 7.62e-08 \\ 
   \midrule
4 & mitochondrial translation & 27/29 & 94/1012 & 5.32e-26 \\ 
   \midrule
5 & response to stimulus & 155/268 & 441/1012 & 2.12e-05 \\ 
   \midrule
6 & positive regulation of double-strand break repair via homologous recombination & 14/18 & 15/1012 & 1.55e-24 \\ 
   \midrule
7 & mitochondrial gene expression & 20/34 & 104/1012 & 4.37e-10 \\ 
   \midrule
8 & transport & 25/43 & 218/1012 & 6.03e-05 \\ 
   \midrule
9 & RNA splicing, via transesterification reactions with bulged adenosine as nucleophile & 24/26 & 57/1012 & 1.88e-28 \\ 
   \midrule
10 & regulation of RNA splicing & 9/16 & 30/1012 & 1.55e-08 \\ 
   \midrule
11 & RNA splicing, via transesterification reactions with bulged adenosine as nucleophile & 19/21 & 57/1012 & 2.26e-21 \\ 
   \midrule
12 & cell communication & 33/63 & 290/1012 & 2.76e-02 \\ 
   \midrule
13 & mitochondrial gene expression & 29/33 & 104/1012 & 1.60e-24 \\ 
   \midrule
14 & proteasome-mediated ubiquitin-dependent protein catabolic process & 24/24 & 59/1012 & 2.08e-30 \\ 
   \midrule
15 & regulation of transcription initiation by RNA polymerase II & 25/63 & 28/1012 & 3.87e-27 \\ 
   \midrule
16 & ribosome biogenesis & 34/51 & 96/1012 & 8.63e-23 \\ 
   \midrule
17 & transcription by RNA polymerase III & 9/31 & 16/1012 & 3.46e-08 \\ 
   \midrule
18 & RNA biosynthetic process & 31/37 & 351/1012 & 3.29e-07 \\ 
   \midrule
19 & cytoplasmic translation & 36/44 & 55/1012 & 9.79e-42 \\ 
   \midrule
20 & NADH dehydrogenase complex assembly & 17/18 & 19/1012 & 1.26e-31 \\ 
   \midrule
21 & response to growth factor & 17/92 & 44/1012 & 3.63e-05 \\ 
   \midrule
22 & mitochondrial translation & 22/22 & 94/1012 & 1.08e-22 \\ 
   \midrule
23 & endomembrane system organization & 5/6 & 30/1012 & 1.06e-05 \\ 
   \bottomrule
\end{tabular}
\caption{Over-representation analyses using our method.}
\label{Table:ORA result}
\end{table}

\begin{table}[!htbp]
\footnotesize
\centering
\begin{tabular}{lp{7cm}ccc}
  \toprule
Cluster & Description & GeneRatio & BgRatio & Adjusted P-Value \\ 
  \midrule
1 & cytoplasmic translation & 11/17 & 55/1012 & 1.68e-08 \\ 
   \midrule
2 & translational initiation & 4/9 & 17/1012 & 7.47e-04 \\ 
   \midrule
3 & positive regulation of DNA biosynthetic process & 8/34 & 11/1012 & 5.91e-08 \\ 
   \midrule
4 & mitochondrial gene expression & 30/34 & 104/1012 & 2.43e-25 \\ 
   \midrule
5 & multicellular organismal process & 162/310 & 343/1012 & 5.35e-13 \\ 
   \midrule
6 & positive regulation of double-strand break repair via homologous recombination & 14/17 & 15/1012 & 3.42e-25 \\ 
   \midrule
7 & mitochondrial RNA metabolic process & 6/27 & 16/1012 & 4.18e-04 \\ 
   \midrule
8 & transport along microtubule & 9/39 & 18/1012 & 1.27e-06 \\ 
   \midrule
9 & RNA splicing, via transesterification reactions with bulged adenosine as nucleophile & 23/25 & 57/1012 & 3.50e-27 \\ 
   \midrule
10 & regulation of RNA splicing & 9/9 & 30/1012 & 6.57e-13 \\ 
   \midrule
11 & RNA splicing, via transesterification reactions with bulged adenosine as nucleophile & 19/23 & 57/1012 & 1.03e-19 \\ 
   \midrule
12 & G protein-coupled receptor signaling pathway & 9/36 & 27/1012 & 5.30e-05 \\ 
   \midrule
13 & mitochondrial translation & 26/30 & 94/1012 & 7.49e-23 \\ 
   \midrule
14 & proteasome-mediated ubiquitin-dependent protein catabolic process & 19/19 & 59/1012 & 3.10e-23 \\ 
   \midrule
15 & regulation of DNA-templated transcription initiation & 27/105 & 29/1012 & 1.49e-23 \\ 
   \midrule
16 & ribosome biogenesis & 31/59 & 96/1012 & 6.07e-16 \\ 
   \midrule
17 & transcription by RNA polymerase III & 10/33 & 16/1012 & 1.14e-09 \\ 
   \midrule
18 & rRNA processing & 12/33 & 63/1012 & 7.94e-05 \\ 
   \midrule
19 & cytoplasmic translation & 21/31 & 55/1012 & 5.93e-19 \\ 
   \midrule
20 & NADH dehydrogenase complex assembly & 17/20 & 19/1012 & 7.96e-30 \\ 
   \midrule
21 & mitotic cell cycle phase transition & 12/49 & 37/1012 & 2.33e-05 \\ 
   \midrule
22 & mitochondrial translation & 28/28 & 94/1012 & 1.31e-29 \\ 
   \midrule
23 & endomembrane system organization & 5/6 & 30/1012 & 1.06e-05 \\ 
   \bottomrule
\end{tabular}
\caption{Over-representation analyses using PPL method.}
\label{Table:ORA result-PPL}
\end{table}

\end{document}